\documentclass[twocolumn,showpacs,preprintnumbers,amsmath,amssymb]{revtex4}

\usepackage{graphicx}
\usepackage{dcolumn}
\usepackage{bm}
\usepackage{epsfig}
\usepackage{bm}

\begin{document}

\title{Cavity phenomena in mesas of cuprate high-$T_c$ superconductors under voltage bias}

\author{Xiao Hu\(^{1,2,3}\) and Shizeng Lin\(^{1,2}\)}

\affiliation{\(^{1}\)WPI Center for Materials Nanoarchitectonics,
 National Institute for Materials Science, Tsukuba 305-0044, Japan\\
\(^{2}\)Graduate School of Pure and Applied Sciences, University of
Tsukuba, Tsukuba 305-8571, Japan\\
\(^{3}\)Japan Science and Technology Agency, 4-1-8 Honcho,
Kawaguchi, Saitama 332-0012, Japan}
\date{\today}

\begin{abstract}
Modeling a single crystal of cuprate high-$T_c$ superconductor, such
as $\rm{Bi_2Sr_2CaCu_2O_{8+\delta}}$, as a stack of intrinsic
Josephson junctions, we formulate explicitly the cavity phenomenon
of plasma oscillations and electromagnetic (EM) waves in mesas of
cylindrical and annular shapes. When the mesa thickness is small
compared with the EM wavelength, the boundary condition for the
inductively coupled sine-Gordon equations is the Neumann-type one to
a good approximation, addressed first theoretically and verified in
a recent experiment. This renders the superconductor mesa a cavity.
Biasing a dc voltage in the $c$ direction, a state with $\pm\pi$
kinks in the superconductivity phase difference piled up
alternatively along the $c$ axis is stabilized. The $\pm\pi$ phase
kinks provide inter-lock between superconductivity phases in
adjacent junctions, taking the advantage of huge inductive couplings
inherent in the cuprate superconductors, which establishes the
coherence across the whole system of more than $\sim 600$ junctions.
They also permit a strong coupling between the lateral cavity mode
and the transverse Josephson plasma, enhance the plasma oscillation
significantly at the cavity modes which radiates EM waves in the
terahertz band when the lateral size of mesa is set to tens of
micrometers. In order to overcome the heating effect, we propose to
use annular geometry. The dependence of frequency on the aspect
ratio is analyzed, which reveals that the shape tailor is quite
promising for improving the present technique of terahertz
excitation. The annular geometry may be developed as a waveguide
resonator, mimic the fiber lasers for visible lights.

\end{abstract}


\maketitle

\section{Introduction}
The Josephson effect provides a unique principle to excite
high-frequency electromagnetic (EM) waves \cite{Josephson,Barone}.
Much effort has been made to stimulate powerful radiations, first
using artificial Josephson junction arrays
\cite{Yanson65,Dayem66,Zimmerma66,Pedersen76,Finnegan72,Jain84,Durala99,Barbara99},
and later on Josephson junctions inherent in cuprate high-$T_c$
superconductors of layered structures
\cite{Kleiner92,Sakai93,Tachiki94,Koyama95,Shafranjuk99,Kume99,Kleiner00,Iguchi00,Batov06,BulaevskiiPRL06,
BulaevskiiJSNM06,BulaevskiiPRL07,HJLeePRL07,KoshelevPRB08,LinHuPRB08}.
The latter have obvious advantages, since the junctions are
homogeneous at the atomic scale guaranteed by the high quality of
single crystals, and the superconductivity gap is large, typically
of tens of meV, which in principle permits the frequency cover the
whole range of the terahertz (THz) band, a very useful regime of EM
waves still lacking of compact solid-state generators
\cite{Ferguson02,Tonouchi07}.

An experimental breakthrough was achieved in 2007 \cite{Ozyuzer07}.
Clear evidences have been obtained that coherent terahertz
radiations from side edges of a thin rectangular mesa were realized
by biasing a dc voltage in the $c$ axis of the
$\rm{Bi_2Sr_2CaCu_2O_{8+\delta}}$ (BSCCO) single crystal; the
frequency of the EM wave and the voltage where the radiation occurs
obey the ac Josephson relation, and the frequency coincides with one
of the cavity modes determined by the lateral size of mesa
\cite{Ozyuzer07}.

This discovery is expected to leave significant and long-standing
impacts. It has the potential to open a practical way to develop a
source of frequency tunable EM waves based on superconductivity and
fill the so-called THz-gap. It also shines light on new directions
of making use of the phase of superconductivity. Its importance
would be better appreciated if one notices that up to now the usage
of superconductivity phase variable is still limited to the SQUID.

The discovery immediately raises many interesting questions, such as
why the Josephson plasma oscillation, namely the coherent tunneling
of Cooper pairs forth and back between adjacent CuO layers driven by
a $c$-axis voltage can radiate strong {\it transverse} radiations,
in absence of Josephson vortices induced by an applied magnetic
field; how it becomes possible to synchronize the superconductivity
phase variables of $\sim600$ junctions, and so on. These quests
challenge our knowledge on superconductivity as well as nonlinear
phenomena.

 It is formulated explicitly in
Ref.~\cite{BulaevskiiPRL07} that there is a significant mismatch in
impedance $|Z_{\rm out}|/|Z_{\rm in}|\sim \lambda_{\rm EM}/L_z$ at
the edge of a superconductor mesa when the thickness of the mesa
$L_z$ is small compared with the EM wavelenght $\lambda_{\rm EM}$,
which was known for a single junction as a limiting case
\cite{Langenberg65}. This makes the tangential component of
oscillating magnetic field at the mesa edge extremely small compared
with the electric one, in sharp contrast to the case of EM plane
waves. Actually, this relation is a general property shared by a
normal capacitor with small electrode separation. Particularly for
thin mesas of superconductor, the vanishingly small tangential
component of magnetic field gives the Neumann-type boundary
condition of superconductivity phase difference across junctions,
namely the spatial derivative of the phase difference normal to the
edge should be zero (in absence of an applied magnetic field).

With this Neumann boundary condition, a new dynamic state of
superconductivity phase difference has been found in a stack of
Josephson junctions under a dc voltage bias, which is characterized
by static $\pm\pi$ (actually $\pm (2m+1)\pi$ with $m$ integer) phase
kinks localized at the mesa center and piled up alternatively along
the $c$ axis
\cite{LinHuPRL08,HuLinPRB08,KoshelevPRB08_2,LinHuPRB09}. In this
$\pi$ kink state, an inter-lock between superconductivity phases in
neighboring junctions appears, which establishes the coherence
across the whole system of more than $\sim 600$ junctions. The $\pi$
phase kink makes the lateral cavity modes and the transverse
Josephson plasma couple strongly to each other, permits large dc
supercurrent flow into the junctions and generates strong EM
radiations from the sides of mesa.

The $\pi$ kink state requests strong inductive couplings between
adjacent superconducting layers \cite{LinHuPRL08}, which is
guaranteed in the cuprates high-$T_c$ superconductor BSCCO. Another
condition is on the thickness of the superconductor mesa: it should
be much smaller than the EM wavelength on one hand, which renders
the superconductor mesa a cavity, and on the other hand, it should
be thick enough to avoid large surface effects. The samples of
$L_z\simeq 1\mu$m adopted in the experiment \cite{Ozyuzer07} satisfy
these requests well. The $\pi$ kink state is stable against other
distortions, such as thermal fluctuations, inhomogeneity of physical
parameters, and so on.

The present work is motivated by two recent experiments on THz
radiations from thin mesas of BSCCO single crystals. The first one
is a detection of THz radiations from a cylindrical mesa
\cite{Kadowaki09}, after the theoretical proposal \cite{HuLinPRB08}
for testifying the boundary condition of the system. The second one
is an observation on EM standing waves in a rectangular mesa
\cite{WangPRL09}, an approach being able to provide direct evidence
for cavity resonance.

 The remaining part of the present paper is organized as
follows. In Sec.~II, we discuss the right boundary condition for
thin mesas from a theoretical point of view. Then we point out that
the result of the recent experiment using a cylindrical mesa is to
be understood in accordance with this boundary condition. In
Sec.~III, the radius dependence of the radiation frequency for first
several modes is presented, accompanied with the spatial
distributions of the $\pi$ kink in superconductivity phase
difference, supercurrent, as well as the EM standing wave.
Section~IV is devoted to a new proposal of using annular mesa, which
can reduce the Joule heating and enhance the leakage of heat as
compared with the cylindrical one. By checking the frequency
variation upon tuning aspect ratio, it is displayed that tailoring
the mesa geometry may be a prosperous way to improve the present
technique for THz stimulation. Finally summary and perspectives are
given in Sec.~V.

\section{Boundary condition and cylindrical mesa}

\subsection{Basic equation and general solution}

The inductively coupled sine-Gordon equations for the
gauge-invariant phase differences in a stack of Josephson junctions
with a dc current bias and dissipations are given in the
dimensionless form as \cite{LinHuPRL08,HuLinPRB08,LinHuPRB09}

\begin{equation}
\Delta P_l=(1-\zeta \Delta^{(2)}) (\sin P_l+\beta\partial_t P_l
+\partial_t^2P_l-J_{\rm{ext}}),
\label{csg}
\end{equation}

\noindent under the Neumann boundary condition in the later
directions $\partial_n P_l=0$;
$\zeta\equiv\lambda_{ab}^2/(s+D)^2\sim 10^5$ is the inductive
coupling and
$\beta\equiv4\pi\sigma_c\lambda_c/c\sqrt{\varepsilon}=0.02$ is the
$c$-axis conductivity; $\Delta $ the Laplace operator in lateral
directions, and $\Delta^{(2)}Q_l=Q_{l+1}+Q_{l-1}-2Q_{l}$ the second
difference operator along the $c$ axis. The lateral space is scaled
by $\lambda_c$ and time by $c/\lambda_c\sqrt{\varepsilon}$. For more
details of definitions see Refs.
\cite{LinHuPRL08,HuLinPRB08,LinHuPRB09}.

The experimental observation of the cavity relation of radiation
frequency \cite{Ozyuzer07,KadowakiPhysicaC08} indicates that
standing waves of plasma oscillation has been built in the cavity
formed by the mesa, which in turn implies that the oscillating part
of the phase difference satisfies the Laplace equation
\cite{LinHuPRL08,HuLinPRB08,LinHuPRB09}. The plasma oscillation
should be uniform since the observed radiations are coherent, known
as a superradiation \cite{Ozyuzer07}. Without losing generality, the
solution to Eq.~(\ref{csg}) can be given by

\begin{equation}
P_l(\bm{r},t)=\omega t + \Tilde{P}(\bm{r},t) + f_lP^s(\bm{r}),
\label{phase}
\end{equation}

\noindent where the first term accounts for the finite dc bias
voltage, and the second for plasma oscillation

\begin{equation}
\Tilde{P}(\bm{r},t)=A g(\bm{r})\sin(\omega t+\varphi)
\label{plasma}
\end{equation}

\noindent with  $A$ the amplitude, $g(\bm{r})$ an eigenfunction of
the Laplace equation with the Neumann boundary condition for the
corresponding geometry, and the frequency given by the voltage
following the ac Josephson relation; the third term carries the
inter-junction coupling via the $l$ dependence.

The general form of Eq.~(\ref{phase}) describes a wealth of
solutions even giving the constraint imposed by available
experimental results. However, it is easy to see
\cite{LinHuPRL08,HuLinPRB08,LinHuPRB09} that the sequences
$f_l=(-1)^l$ or $f_l=(-1)^{[l/2]}$ with period 2 or 4 layers
respectively diagonalize Eq.~(\ref{csg}) due to the property of the
difference operator $\Delta^{(2)}$. In simulations, these two states
appear frequently and are stable \cite{LinHuPRL08,LinHuPRB09}. Here
we focus the attention to these two cases.

The static phase $P^s(\bm{r})$ should then satisfy the differential
equation

\begin{equation}
\Delta P^s=q\zeta \cos\varphi J_1(A g(\bm{r}))\sin P^s,
\label{phasekink}
\end{equation}

\noindent with the same boundary condition of the total phase
difference, where $q=4$ and 2 for the $c$-axis sequence of period 2
and 4 respectively.

The static phase factor $P^s(\bm{r})$, amplitude $A$, phase shift in
plasma oscillation $\varphi$ and total current $J_{\rm ext}$ can be
solved for a given voltage (or equivalently $\omega$)
\cite{LinHuPRL08,HuLinPRB08}. The nontrivial solution $P^s(\bm{r})$
presumes the $\pi$ phase kink
\cite{LinHuPRL08,HuLinPRB08,KoshelevPRB08_2,LinHuPRB09}. The static
$c$-axis supercurrent which contributes to the net current flowing
into the system in the $c$ direction is given by

\begin{equation}
J_s(\bm{r})=\sin\varphi J_1\bigl(A g(\bm{r})\bigr)\cos
\bigl(P^s(\bm{r})\bigr). \label{Js}
\end{equation}

\noindent We point out that the claim in Ref.~\cite{TachikiPRL09} on
a very nonlinear \emph{IV} characteristics in the absence of $\pi$
kinks is not compatible with Eq.~(\ref{Js}) and needs further check.

The trivial vacua solution $P^s=0$ is dropped here, which gives a
linear \emph{IV} and a broad voltage regime with weak radiations,
inconsistent with experiments \cite{Ozyuzer07,KadowakiPhysicaC08}.
See Ref.~\cite{MatsumotoPhysicaC08,LinHuPRB09} for more discussions
on this vacua state.

It is noticed that only the fundamental frequency is taken in
Eq.~(\ref{plasma}) since higher harmonics are small especially in
cylindrical and annular geometry, where the higher harmonic
frequencies appearing in the expansion of the sine function of the
dc Josephson relation are not the cavity frequency.

The oscillating electric and magnetic fields inside the junctions
are given by the plasma oscillation term of the phase difference in
dimensionless forms as

\begin{equation}\label{E}
E^z(\bm{r},t)=\frac{\partial \Tilde{P}}{\partial t},
\end{equation}

\noindent with $E^x, E^y\simeq 0$ to a good approximation in
cuprates \cite{LinHuPRL08}, and

\begin{equation}\label{B}
\bm{B}(\bm{r},t)=-\bm{\nabla}\times
(\Tilde{P}(\bm{r},t)\hat{\bm{z}}),
\end{equation}

\noindent with $\hat{\bm{z}}$ the unit vector in the $c$ axis. It is
Eq.~(\ref{B}) that enforces the boundary condition for the phase
difference $\partial_n P_l=0$ with $n$ the normal of the sample
edges.

\subsection{Testifying the boundary condition}

In a rectangular mesa, the spatial part of the plasma term of the
lowest cavity mode satisfying the right boundary condition is
$\cos(x\pi/L)$. Unfortunately, $\cos(x\pi/L)$ and $\sin(x\pi/L)$ are
both eigenfunctions of the Laplace equation, derivative to each
other, and give the same wave number (or equivalently frequency in
dimensionless form) $\pi/L$. Therefore, from a pure experimental
point of view, the dependence of frequency on the system size cannot
determine uniquely the mode.

A mesa of cylindrical shape was proposed to testify uniquely the
dynamics of the superconductivity phase difference inside the
junctions \cite{HuLinPRB08}. For the cylinder geometry, the radial
part of an eigenfunction of the Laplace equation is given by the
Bessel function, not their derivatives. The boundary condition for
the cylinder geometry determines uniquely the wave number, and vice
versa, since the zeros of the Bessel functions are different from
the zeros of their derivatives, contrasting to sine and cosine
functions in the rectangle geometry. Therefore, measuring by
experiments the frequency of EM radiation from a cylindrical mesa of
given radius and assigning the mode by the properties of Bessel
functions enable one to identify uniquely the right boundary
condition for the EM waves. It was addressed that, from the boundary
condition suitable for thin mesas, the frequency of radiation should
be given by the zeros of derivatives of Bessel functions
\cite{HuLinPRB08}. A case of the (0,1) mode, which is uniform
azimuthally, has been worked out explicitly, and the frequency is
figured out as $k^{\rm c}_{01}=\chi^{\rm c}_{01}/a$ with the radius
$a$ and $\chi^{\rm c}_{01}=3.8317$ the first zero of derivative of
the Bessel function $J_0(z)$ \cite{Note}.

The recent experiment \cite{Kadowaki09} showed that the frequency of
EM wave radiated from the cylindrical mesa of radius $a\simeq
45\mu$m is $f\simeq 0.474$THz. With the light velocity in the sample
$c'=c/\sqrt{\varepsilon}\simeq 7.16\times 10^7$m/s
($\varepsilon\simeq 17.54$ \cite{Kadowaki08}), the wave number
observed in the experiment \cite{Kadowaki09} is given by $k\simeq
1.87/a$. Since $\chi=1.8412$ is the first zero of derivative of the
Bessel function $J_1(z)$, it is inferred that the mode (1,1) has
been achieved in the experiment. The most important message from the
experiment is that the frequency of EM radiation is determined by
the zero of derivative of the Bessel function, in agreement with the
theoretical analysis \cite{HuLinPRB08}. The right boundary condition
for the cavity therefore is that the tangential component of the
oscillating magnetic field should be zero at the edges as discussed
above, consistent with the analysis in Ref.~\cite{BulaevskiiPRL07}.
The (1,1) cavity mode for the cylinder geometry assumes a lower
frequency than the (0,1) mode due to the positions of zeros of
derivatives of Bessel functions  \cite{Note}, and thus is realized
at a low voltage while the (0,1) mode might be difficult to excite
in experiment where the heating effect is crucial.

\section{Modes for cylindrical mesa}

Knowing the right boundary condition, it is not hard to figure out
the EM standing waves for all the cavity modes. Since their
distribution can be observed directly in experiments
\cite{WangPRL09} and serves as a check of the theory, we map them
out explicitly. The distributions of superconductivity phase
difference and the supercurrent are helpful for understanding the
way how large dc powers are converted to high-frequency transverse
radiations.

\begin{figure}[t]
\epsfysize=8cm \epsfclipoff \fboxsep=0pt
\setlength{\unitlength}{1cm}
\begin{picture}(8,6)(0,0)
\epsfysize=3.0cm \put(0.0,3.0){\epsffile{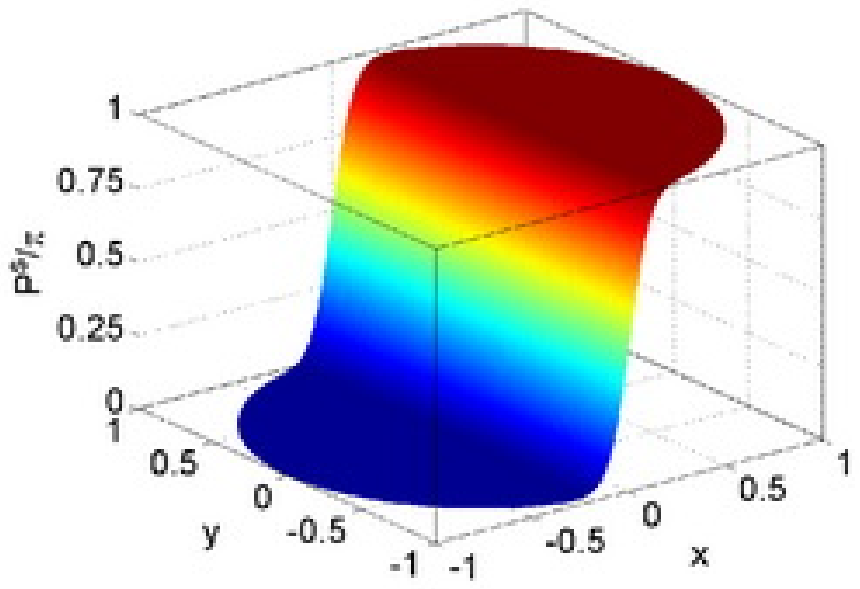}}
\epsfysize=3.0cm \put(4.0,3.0){\epsffile{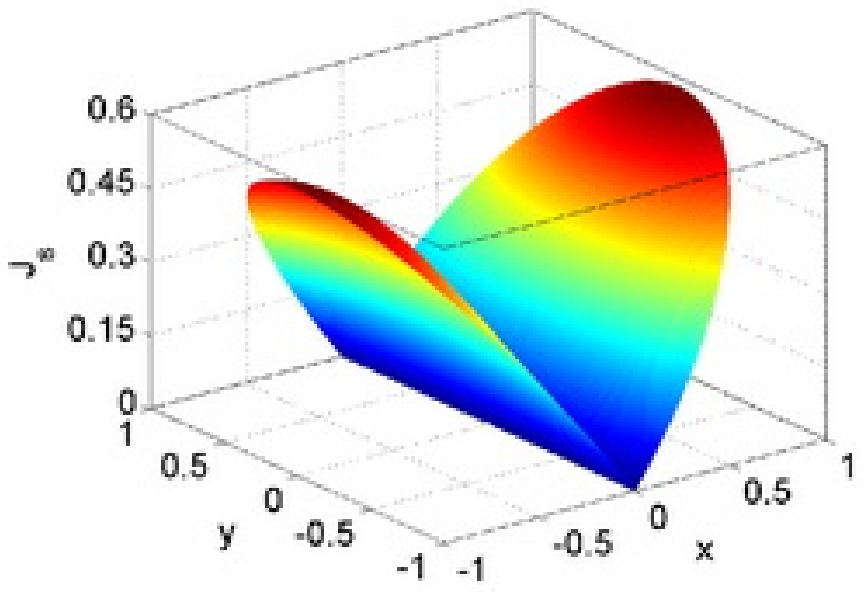}}
\epsfysize=3.0cm \put(0.0,0.0){\epsffile{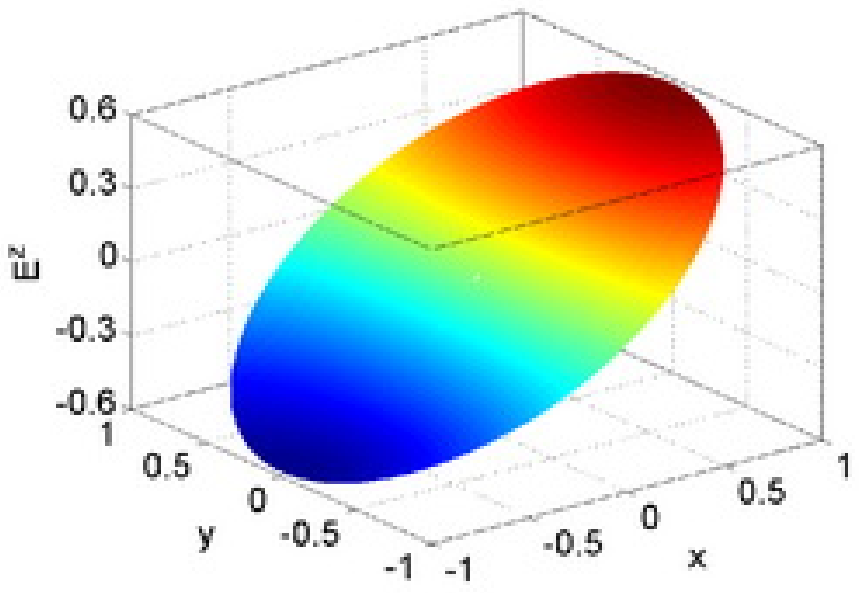}}
\epsfysize=3.0cm \put(4.5,0.0){\epsffile{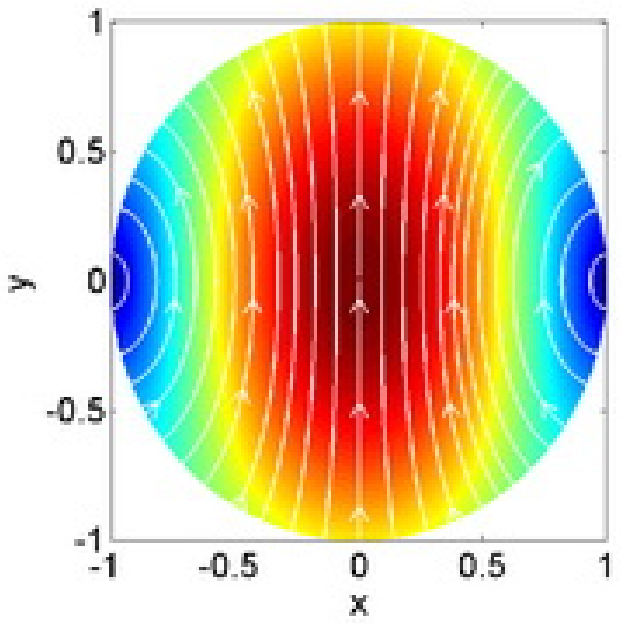}}
\put(0.2,5.8){(a)} \put(4.2,5.8){(b)} \put(0.2,2.8){(c)}
\put(4.2,2.8){(d)}
\end{picture}
\caption{(color online). Spatial distribution of (a) the static
phase term $P^s$, (b) supercurrent, and the standing (c) electric
and (d) magnetic wave for the (1,1) mode of a cylindrical mesa. Here
$\zeta'= qA\zeta\sin\varphi = 5000$ is taken for
Eq.~(\ref{phasekink}) in (a) and the kink is approximated as a step
function in (b) corresponding to $\zeta=\infty$. The lateral
coordinates are normalized by the radius of the cylinder. The
quantities except the phase difference $P^s$ are up to the plasma
amplitude $A$.} \label{cm11}
\end{figure}

\subsection{(1,1) mode for cylindrical mesa}

 All the modes $(m,n)$ for the cylindrical geometry can be classified according
 to the number of symmetry axes in the azimuthal direction, $m$, and the
 number of maxima and minima along the radial direction, $n$, in the spatial part
 of plasma oscillation and electric field (see Eq.~(\ref{E})),
 which all correspond to the zeroes of the derivative of Bessel functions.
 The solution to Eq.~(\ref{csg}) assigned as the (1,1)
mode for the cylinder geometry is given by

\begin{equation}
P_l(\bm{r},t)=\omega t+AJ_1\Bigl(\frac{\chi^{\rm
c}_{11}}{a}\rho\Bigr)\cos\phi \sin(\omega t+\varphi)+
f_{l}P^s(\bm{r}), \label{phasecm11}
\end{equation}

\noindent where the cylindrical coordinates are given by
$\bm{r}=(\rho,\phi)$; $\chi^{\rm c}_{11}=1.8412$ is the first zero
of the derivative of $J_1(z)$. For a cylinder with uniform physical
properties, the eigenfunction for the azimuthal angle $\sin\phi$
degenerates with $\cos\phi$, which, and any linear combination of
them, can be absorbed into the latter by redefining the azimuthal
angle $\phi$, and will not be discussed explicitly.

The wave number, and equivalently frequency ($k=f$ in dimensionless
form and $k=c'f$ with units), of this mode is given by $k^{\rm
c}_{11}=\chi^{\rm c}_{11}/a=1.8412/a$, which agrees well with the
experiment \cite{Kadowaki09} as discussed in the previous section.

The distribution of the static phase $P^s$, supercurrent and the
spatial part of the oscillating electric and magnetic fields are
displayed in Fig.~\ref{cm11}. The $\pi$ phase kink runs along the
diameter at the direction $\phi=\pm\pi/2$, and the phase difference
saturates to 0 and $\pi$ at the left and right part of the cylinder
(Fig.~\ref{cm11}(a)).

Associated with the static phase kink, the oscillating magnetic
field assumes the maximal absolute value along the diameter at
$\phi=\pm\pi/2$, and decreases to zero at the left- and right-most
parts of the cylinder (Fig.~\ref{cm11}(d)). The azimuthal component
of the magnetic field is always zero at the edge of cylinder, as
imposed by the boundary condition.  The electric field takes its
maximal absolute value at the left- and right-most parts, while is
reduced to zero along the diameter at $\phi=\pm\pi/2$
(Fig.~\ref{cm11}(c)).

Comparing the distributions of the $\pi$ phase kink and the electric
and magnetic field, it is found that the (1,1) mode for the cylinder
geometry corresponds to the (1,0) mode for the rectangle geometry
\cite{HuLinPRB08}.

The maximal value of the supercurrent takes place at the left- and
right-most part of the cylinder for the present mode. The factor
$\cos \bigl(P^s(\bm{r})\bigr)$ from the $\pi$ phase kink
(Fig.~\ref{cm11}(a)) renders the supercurrent associated with the
cavity mode $J_1\bigl(A g(\bm{r})\bigr)$ positive over the cylinder
(Fig.~\ref{cm11}(b)), which permits large bias current when the
plasma amplitude $A$ is enhanced at the cavity resonance.

\begin{figure}[t]
\epsfysize=8cm \epsfclipoff \fboxsep=0pt
\setlength{\unitlength}{1cm}
\begin{picture}(8,6)(0,0)
\epsfysize=3.0cm \put(0.0,3.0){\epsffile{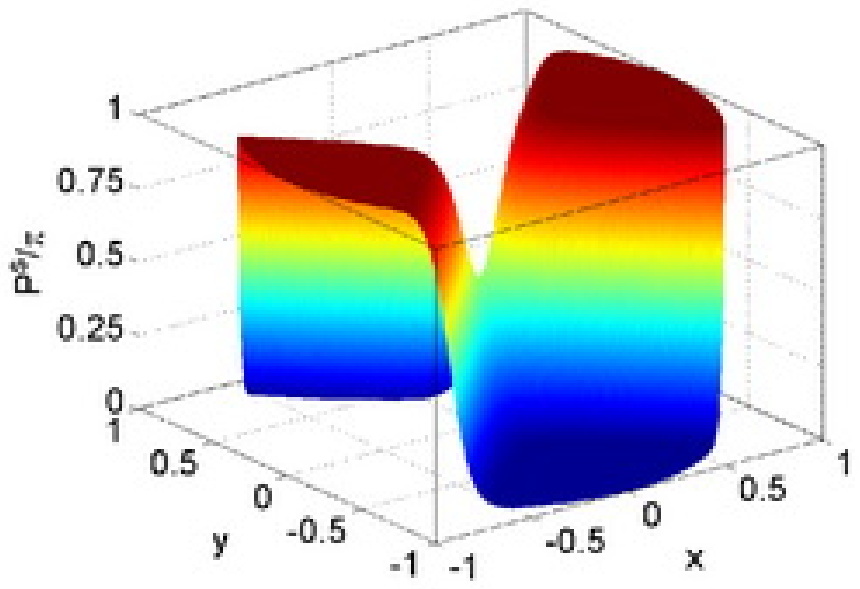}}
\epsfysize=3.0cm \put(4.0,3.0){\epsffile{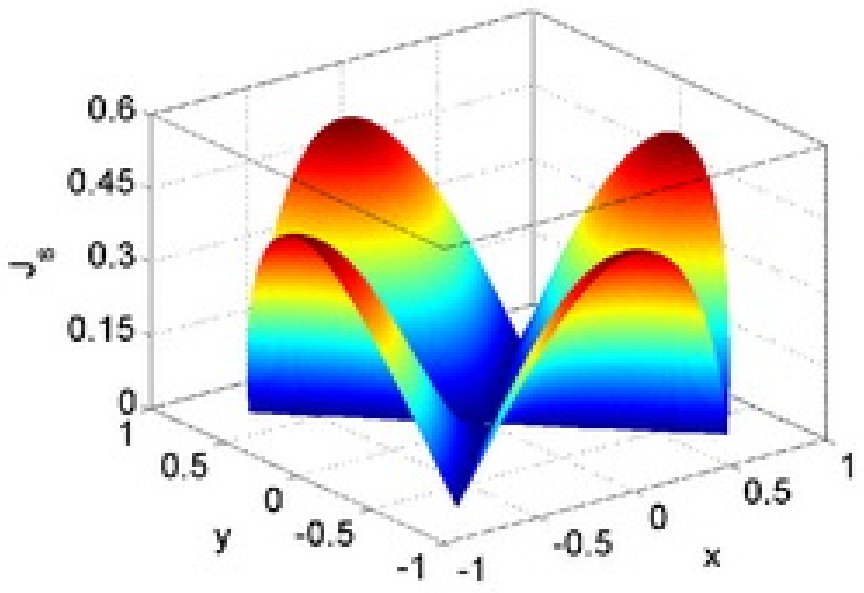}}
\epsfysize=3.0cm \put(0.0,0.0){\epsffile{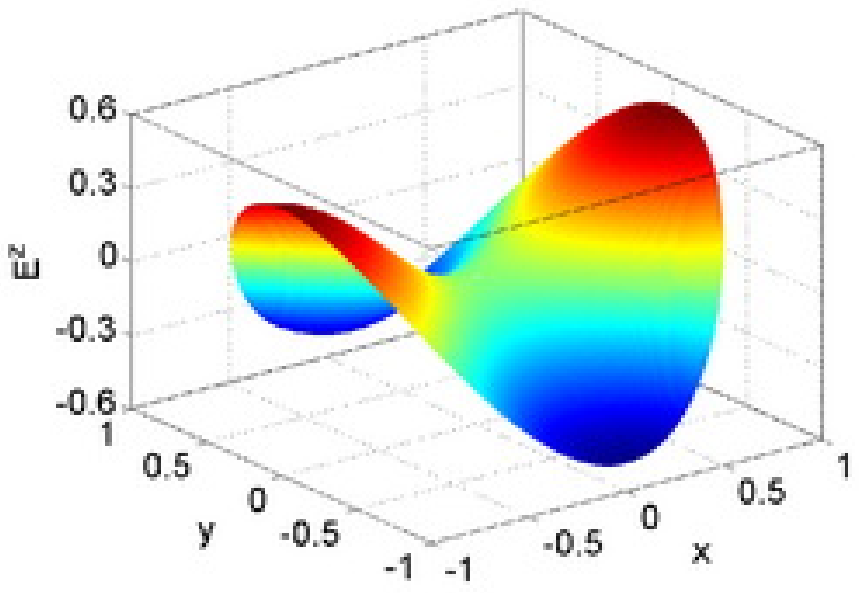}}
\epsfysize=3.0cm \put(4.5,0.0){\epsffile{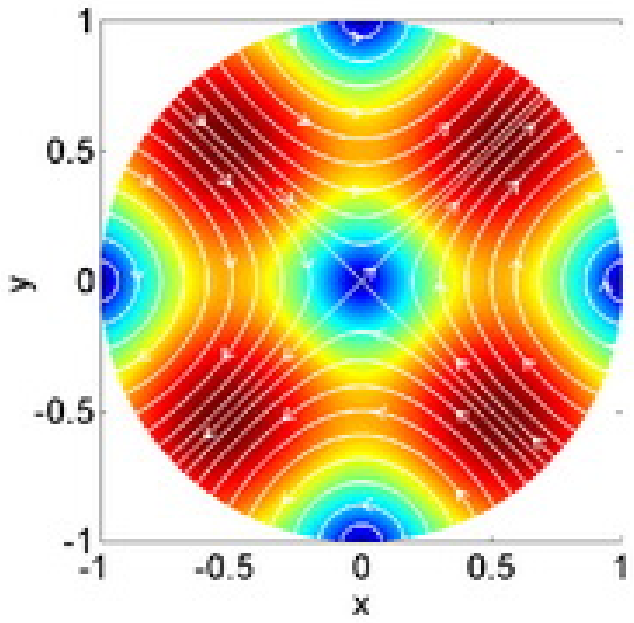}}
\put(0.2,5.8){(a)} \put(4.2,5.8){(b)} \put(0.2,2.8){(c)}
\put(4.2,2.8){(d)}
\end{picture}
\caption{(color online). Same as Fig.~\ref{cm11} for the (2,1) mode
of
  a cylindrical mesa.} \label{cm21}
\end{figure}

\subsection{(2,1) mode for cylindrical mesa}

The (2,1) mode in the cylindrical mesa is given by

\begin{equation}\label{phasecm21}
P_l(\bm{r},t)=\omega t+AJ_2\Bigl(\frac{\chi^{\rm
c}_{21}}{a}\rho\Bigr)\cos(2\phi) \sin(\omega t+\varphi)+
f_{l}P^s(\bm{r}),
\end{equation}

\noindent with $\chi^{\rm c}_{21}=3.0542$ the first zero of the
derivative of $J_2(z)$. The patterns for the (2,1) mode are
displayed in Fig.~\ref{cm21}. There are two pairs of $\pm\pi$ kinks
in the azimuthal direction (Fig.~\ref{cm11}(a)), which increase the
frequency to $k^{\rm c}_{21}=3.0542/a$ higher than the (1,1) mode
where the $\pi$ kink running along a diameter (Fig.~\ref{am11}(a))
equivalent to a pair of $\pm\pi$ kinks in the azimuthal direction.

The magnetic field penetrates into the cylinder along two directions
(Fig.~\ref{cm21}(d)), $\phi=3\pi/4$ and $\phi=-\pi/4$, and flows
away along the two orthogonal directions, $\phi=\pi/4$ and
$\phi=-3\pi/4$, where the absolute value of magnetic field assumes
its maximum. The electric field and the supercurrent become maximal
at the directions of multiples of $\phi=\pi$'s. The (2,1) mode for
the cylinder geometry corresponds to the (1,1) mode for the
rectangle geometry \cite{HuLinPRB08}.

\begin{figure}[t]
\epsfysize=8cm \epsfclipoff \fboxsep=0pt
\setlength{\unitlength}{1cm}
\begin{picture}(8,6)(0,0)
\epsfysize=3.0cm \put(0.0,3.0){\epsffile{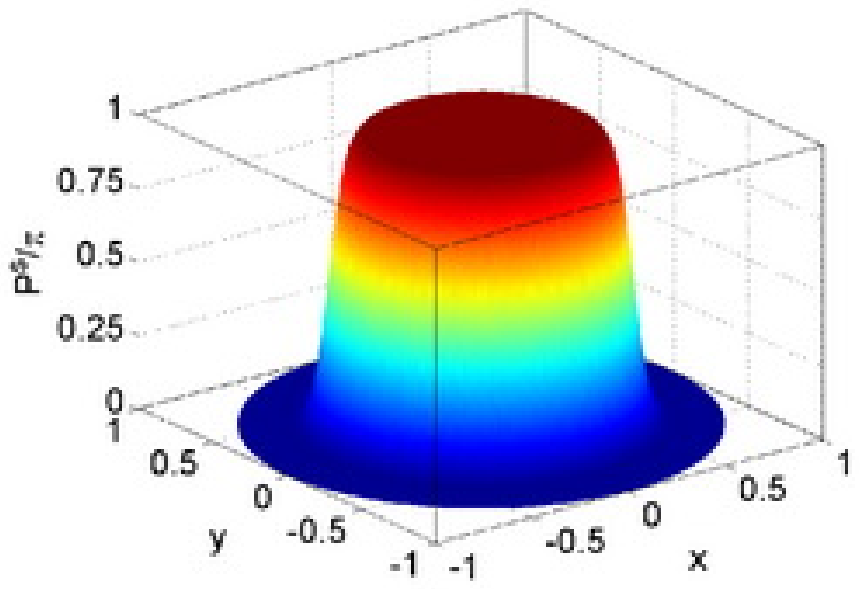}}
\epsfysize=3.0cm \put(4.0,3.0){\epsffile{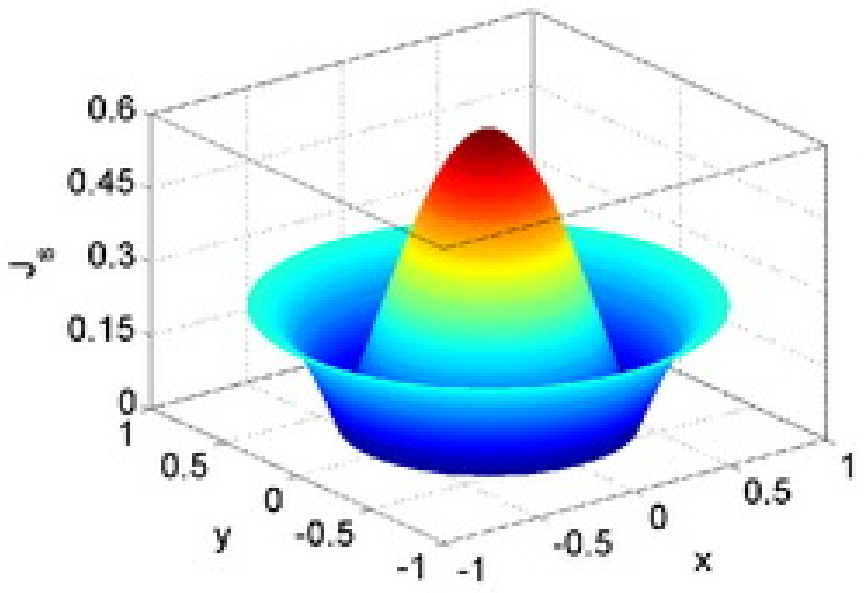}}
\epsfysize=3.0cm \put(0.0,0.0){\epsffile{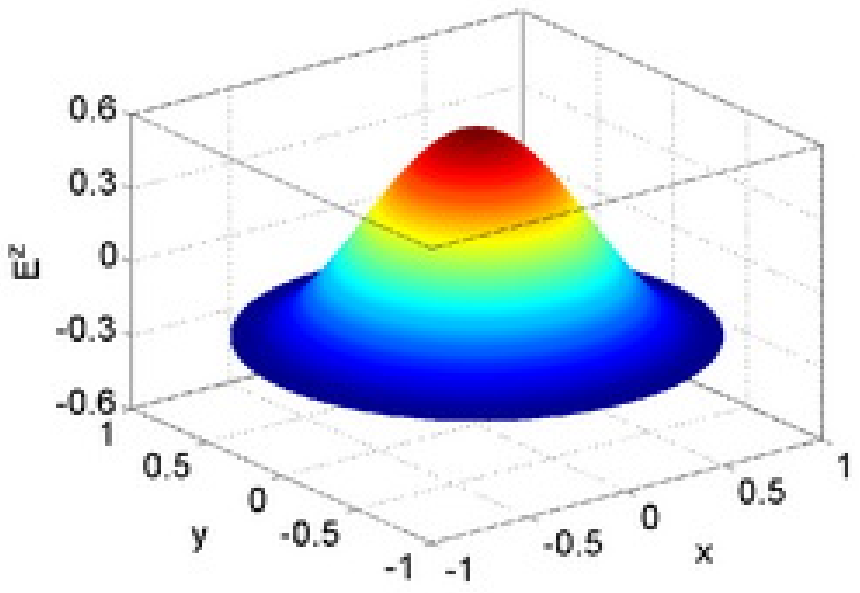}}
\epsfysize=3.0cm \put(4.5,0.0){\epsffile{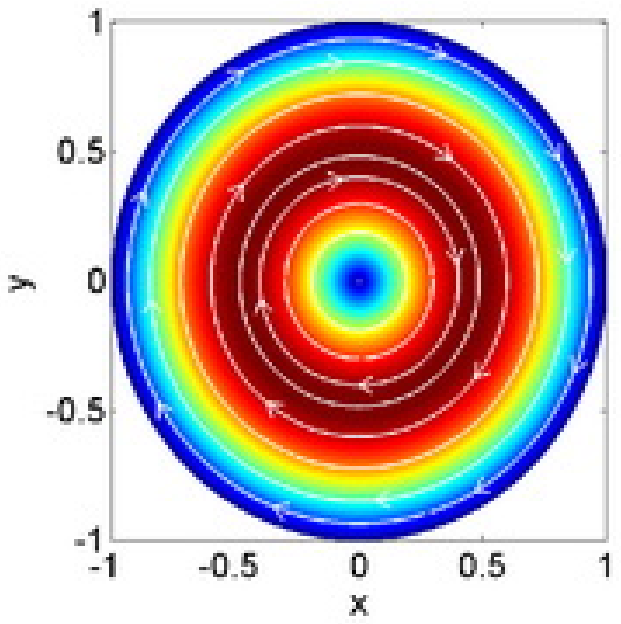}}
\put(0.2,5.8){(a)} \put(4.2,5.8){(b)} \put(0.2,2.8){(c)}
\put(4.2,2.8){(d)}
\end{picture}
\caption{(color online). Same as Fig.~\ref{cm11} for the (0,1) mode
of
  a cylindrical mesa.} \label{cm01}
\end{figure}

\subsection{(0,1) mode for cylindrical mesa}

For comparison, we display in Fig.~\ref{cm01} the patterns for the
(0,1) mode for cylinder geometry which has been discussed in
Ref.~\cite{HuLinPRB08}. Since this mode given by

\begin{equation}
P_l(\bm{r},t)=\omega t+AJ_0\Bigl(\frac{\chi^{\rm
c}_{01}}{a}\rho\Bigr) \sin(\omega t+\varphi)+ f_{l}P^s(\bm{r}),
\label{phasecm01}
\end{equation}

\noindent with $\chi^{\rm c}_{01}=3.8317$ the first zero of the
derivative of $J_0(z)$, is uniform azimuthally, the $\pi$ phase kink
has to be compressed into the radial direction (Fig.~\ref{cm01}(a)).
This makes the spatial variation of the magnetic field steep
(Fig.~\ref{cm01}(d)), and thus the wave number increases to $k^{\rm
c}_{01}=3.8317/a$, which is to be excited at a voltage higher than
the above (1,1) and (2,1) modes. The magnetic field is circular in
this mode (Fig.~\ref{cm01}(d)), and in order to match the boundary
condition at the edge it is suppressed to zero totally. The magnetic
field is also suppressed to zero at the center.

\begin{figure}
\epsfysize=8cm \epsfclipoff \fboxsep=0pt
\setlength{\unitlength}{1cm}
\begin{picture}(8,6)(0,0)
\epsfysize=3.0cm \put(0.0,3.0){\epsffile{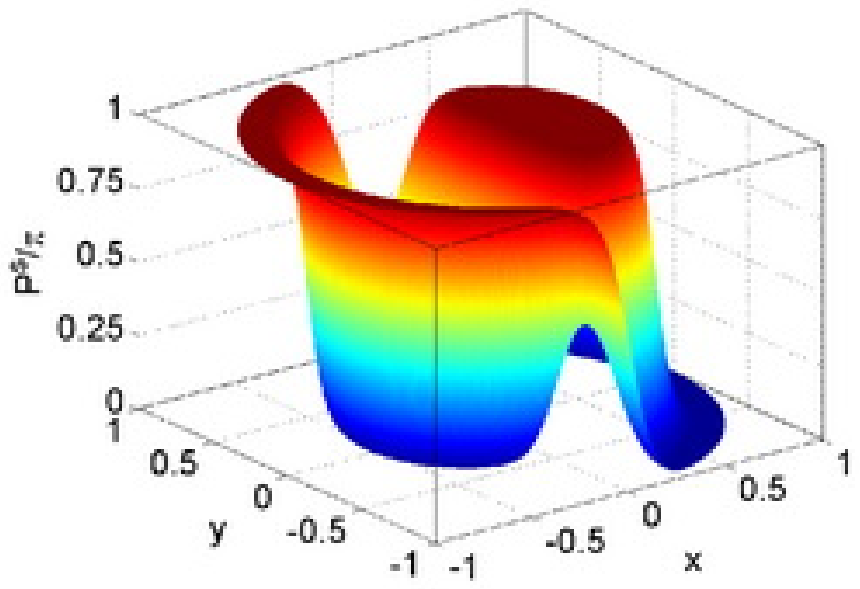}}
\epsfysize=3.0cm \put(4.0,3.0){\epsffile{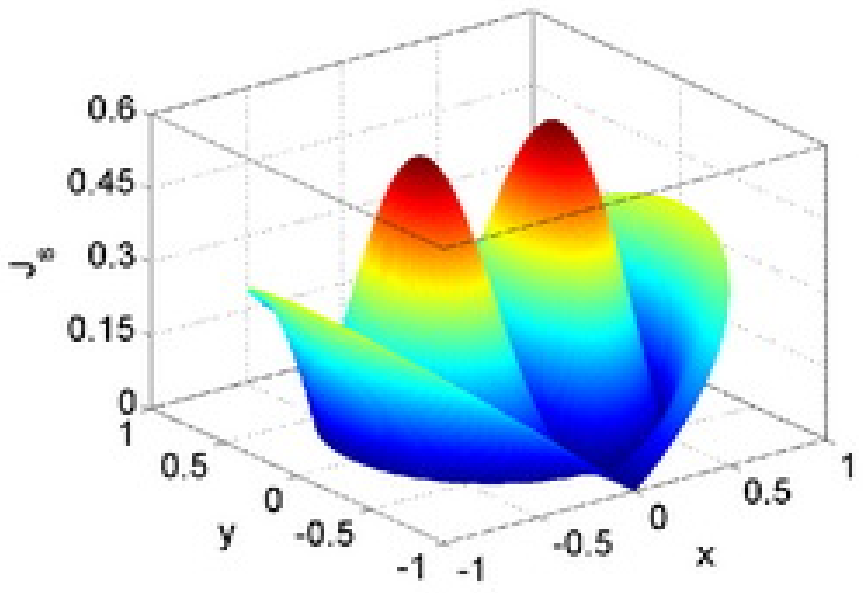}}
\epsfysize=3.0cm \put(0.0,0.0){\epsffile{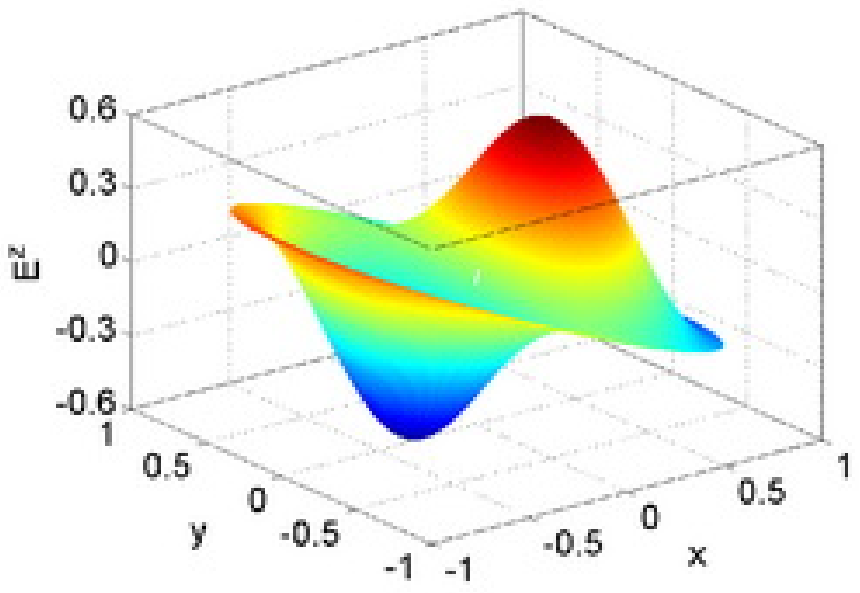}}
\epsfysize=3.0cm \put(4.5,0.0){\epsffile{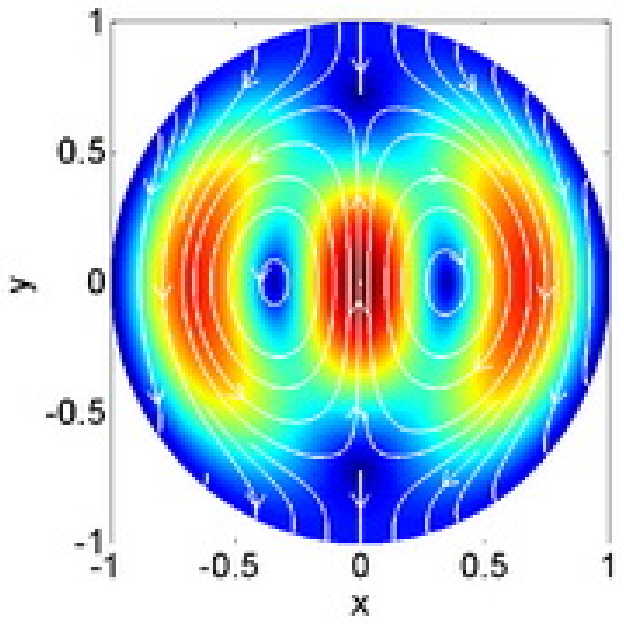}}
\put(0.2,5.8){(a)} \put(4.2,5.8){(b)} \put(0.2,2.8){(c)}
\put(4.2,2.8){(d)}
\end{picture}
\caption{(color online). Same as Fig.~\ref{cm11} for the (1,2) mode
of
  a cylindrical mesa.} \label{cm12}
\end{figure}

\subsection{(1,2) mode for cylindrical mesa}

The $\pi$ phase kink can occur simultaneously in both azimuthal and
radial directions. The lowest mode is the (1,2) mode as shown in
Fig.~\ref{cm12} given by

\begin{equation}
P_l(\bm{r},t)=\omega t+AJ_1\Bigl(\frac{\chi^{\rm
c}_{12}}{a}\rho\Bigr) \cos\phi \sin(\omega t+\varphi)+
f_{l}P^s(\bm{r}), \label{phasecm12}
\end{equation}

\noindent with $\chi^{\rm c}_{12}=5.3314$ the second zero of
derivative of $J_1(z)$.

\subsection{\emph{IV} characteristics for cylindrical mesa}

The \emph{IV} characteristics associated with a cavity mode denoted
by the eigenfunction $g(\bm{r})$ is well approximated by
\cite{LinHuPRL08,HuLinPRB08,KoshelevPRB08_2,LinHuPRB09}

\begin{figure}[b]
\setlength{\unitlength}{1cm}
 \psfig{figure=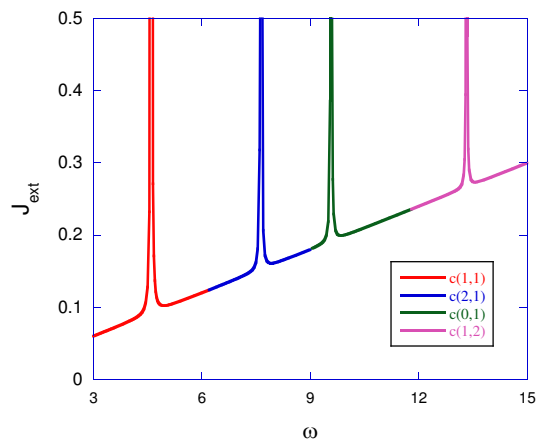,width=7cm}
\caption{(color online). \emph{IV} characteristics for the
cylindrical mesa including the four lowest modes. The dimensionless
voltage, equal to the wave number and frequency, is given by
$\omega=\chi^c/a$ with $\chi^c=1.8412$, 3.0542, 3.8317 and 5.3314
for the (1,1), (2,1), (0,1) and (1,2) mode, respectively. The radius
of cylinder is $a=0.4$ ($a=0.4 \lambda_c$).} \label{ivcm}
\end{figure}

\begin{equation}
J_{\rm ext}=\beta\omega \Biggl[1+\frac{\Bigl(\int_{\Omega}
d\bm{r}g(\bm{r})\cos P^s\Bigr)^2}{2\Omega \int_{\Omega}
d\bm{r}\bigl(g(\bm{r})\bigr)^2}\frac{1}{(\omega^2-k^2)^2+(\beta\omega)^2}
\Biggr],
\label{iv}
\end{equation}

\noindent with $k$ the cavity frequency and $\Omega$ the cross
section of mesa perpendicular to the $c$ axis. The prefactor of the
nonlinear resonating term is 0.379, 0.345, 0.357 and 0.314 for the
(1,1), (2,1), (0,1) and (1,2) modes, respectively. In
Fig.~\ref{ivcm}, we display the \emph{IV} characteristics for a
cylinder with the radius $a=0.4$. A treatment with improvement on
the peak value of current at the resonance \cite{HuLinPRB08} is also
available for the present cylindrical geometry.

\subsection{Radiation pattern for cylindrical mesa}

\begin{figure}
\epsfysize=8cm \epsfclipoff \fboxsep=0pt
\setlength{\unitlength}{1cm}
\begin{picture}(8,6.8)(0,0)
\epsfysize=3.5cm \put(0.0,3.5){\epsffile{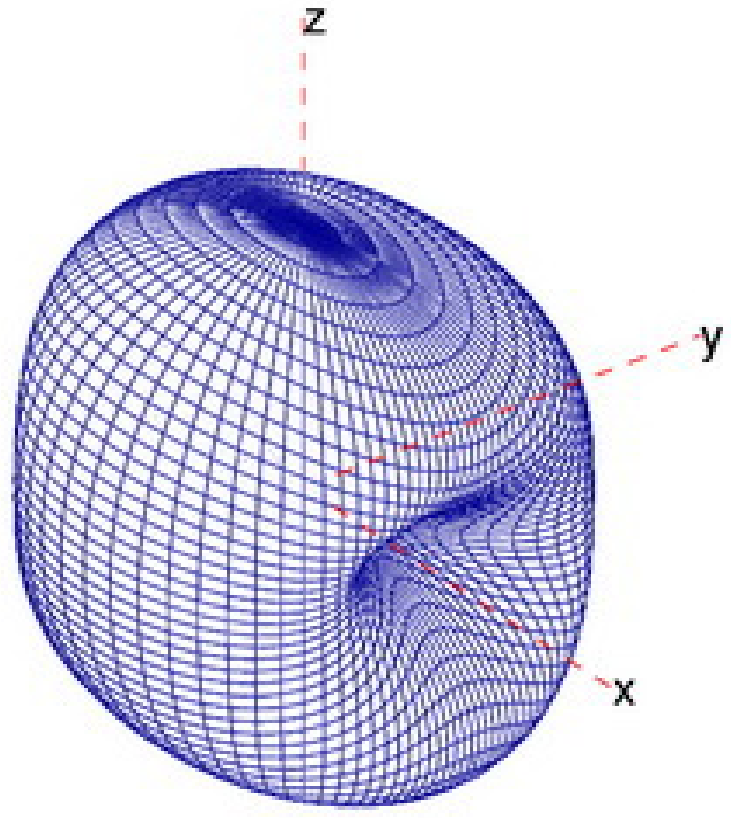}}
\epsfysize=4.0cm \put(4.0,3.2){\epsffile{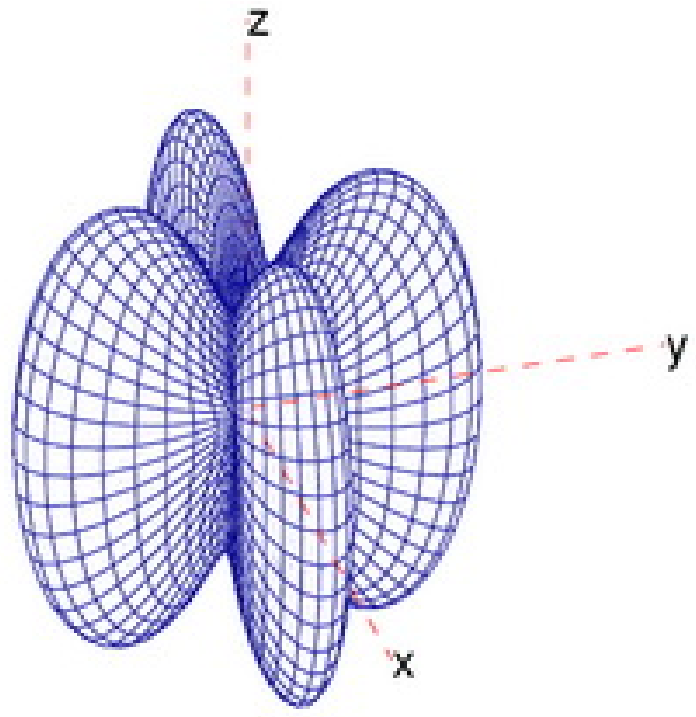}}
\epsfysize=3.5cm \put(0.0,0.0){\epsffile{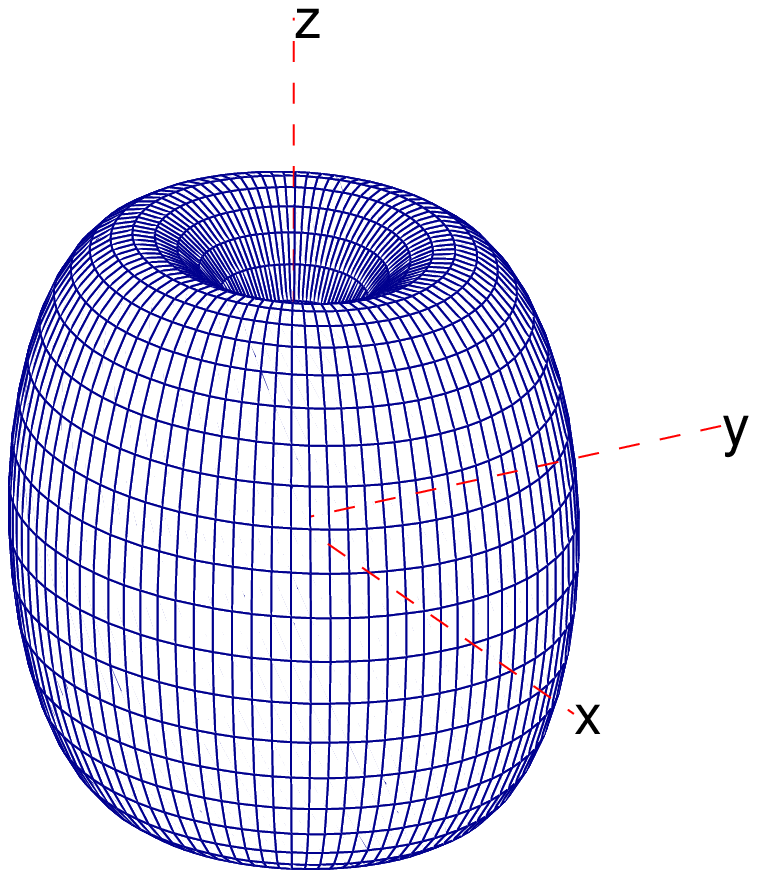}}
\epsfysize=3.5cm \put(4.0,0.0){\epsffile{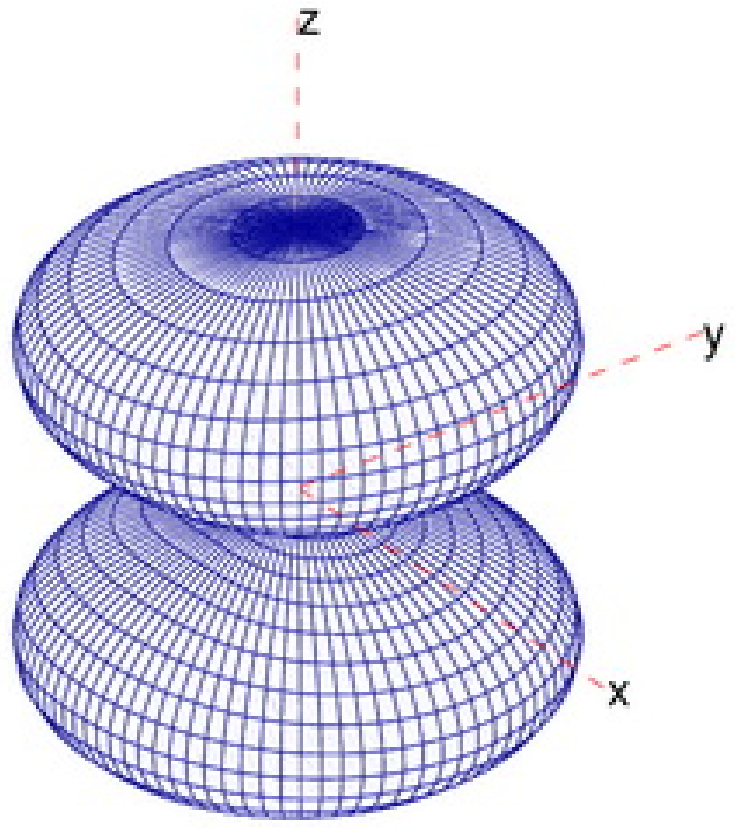}}
\put(0.2,6.6){(a)} \put(4.2,6.6){(b)} \put(0.2,3.1){(c)}
\put(4.2,3.1){(d)}
\end{picture}
\caption{(color online). Radiation patterns for (a) (1,1) mode, (b)
(2,1) mode,
  (c) (0,1) mode, and (d) (1,2) mode of cylindrical mesa at the
  respective resonance frequency. The radius of the mesa is $a=0.4$.}
  \label{patterncm}
\end{figure}

 The radiation pattern of each mode can be computed by
 using the Huygens principle \cite{LinHuPRB09}, and the result is
 displayed in Fig.~\ref{patterncm}.

\section{Annular mesa}

One of the hurdles for the present technique to excite EM waves of
even higher frequency is the heating effect, since the corresponding
higher dc voltage injects large currents into the sample resulting
in severe Joule heating. One way to overcome this effect may be to
dig a hole in the superconductor mesa, rendering for example a
cylindrical one to annular, which reduces the cross section, and
thus the total current and Joule heating. The inner surface of an
annular mesa may help leaking heat generated in the mesa
additionally. Tailoring the shape of superconductor mesa however
will affect the cavity mode, and thus the radiation frequency in a
nontrivial way. In order to check this effect, we need to understand
the cavity phenomenon for the annular geometry.

\vspace{5mm}

\subsection{(1,1) mode for annular mesa}

The (1,1) mode in an annular mesa is given by

\begin{widetext}
\begin{equation}
P_l(\bm{r},t)=\omega t+ \Bigl(A_{\rm J}
J_1(\frac{\chi^{a}_{11}}{a_{\rm o}}\rho) + A_{\rm N}
N_1(\frac{\chi^{a}_{11}}{a_{\rm o}}\rho )\Bigr) \cos\phi \sin(\omega
t+\varphi)+ f_{l}P^s(\rho), \label{phaseam11}
\end{equation}
\end{widetext}

\noindent where $N_1(z)$ is the Bessel function of second kind, and
$a_{\rm o}$($a_{\rm i}$) the outer(inner) radius. The Neumann
boundary condition should be satisfied at both the outer and inner
surfaces, which determines the coefficient $\chi^{a}_{11}$ for a
given radius ratio $a_{\rm i}/a_{\rm o}$:

\begin{equation}
J'_1(\chi^{a}_{11}\frac{a_{\rm i}}{a_{\rm o}})N'_1(\chi^{a}_{11}) -
J'_1(\chi^{a}_{11})N'_1(\chi^{a}_{11}\frac{a_{\rm i}}{a_{\rm o}})=0,
\label{vam11}
\end{equation}

\noindent where $J'_1(z)$ and $N'_1(z)$ are the first derivatives,
and the wave number, or equivalently the frequency, is given by
$k^{a}_{11}=\chi^{a}_{11}/a_{\rm o}$. The ratio of the coefficients
is then determined by

\begin{equation}
A_{\rm N}/A_{\rm J}=-J'_1(\chi^{a}_{11})/N'_1(\chi^{a}_{11}).
\label{Aam11}
\end{equation}

\begin{figure}
\epsfysize=8cm \epsfclipoff \fboxsep=0pt
\setlength{\unitlength}{1cm}
\begin{picture}(8,6)(0,0)
\epsfysize=3.0cm \put(0.0,3.0){\epsffile{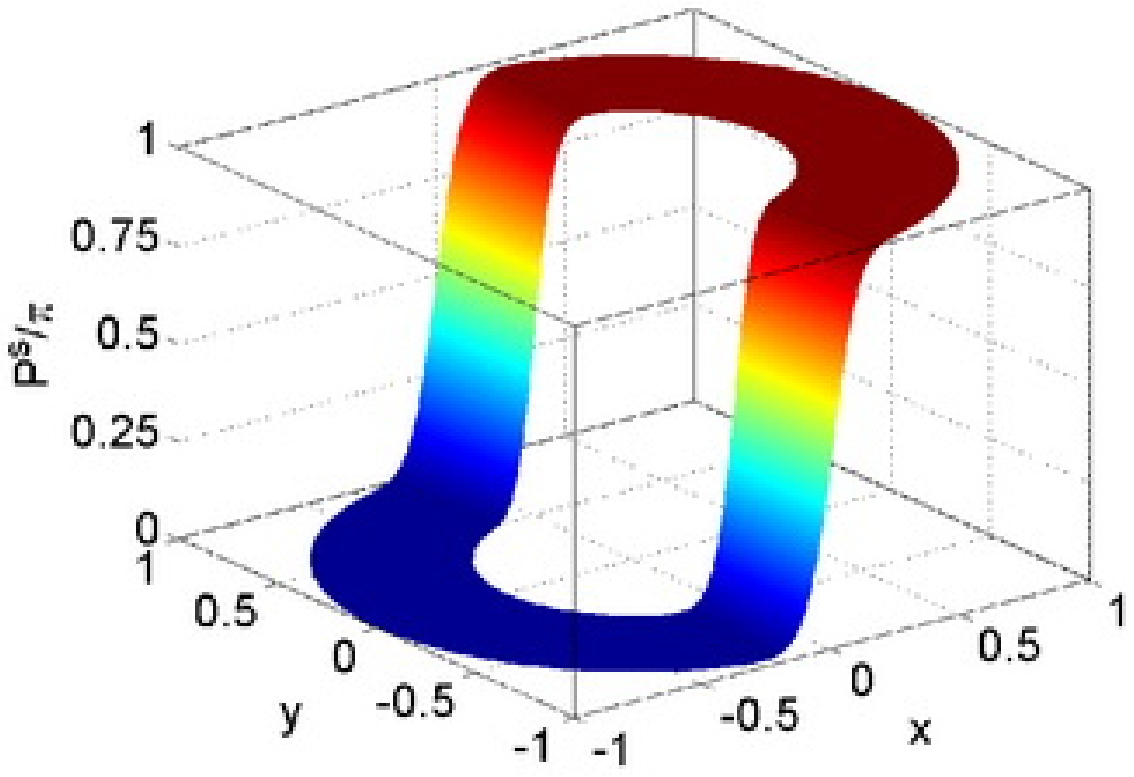}}
\epsfysize=3.0cm \put(4.0,3.0){\epsffile{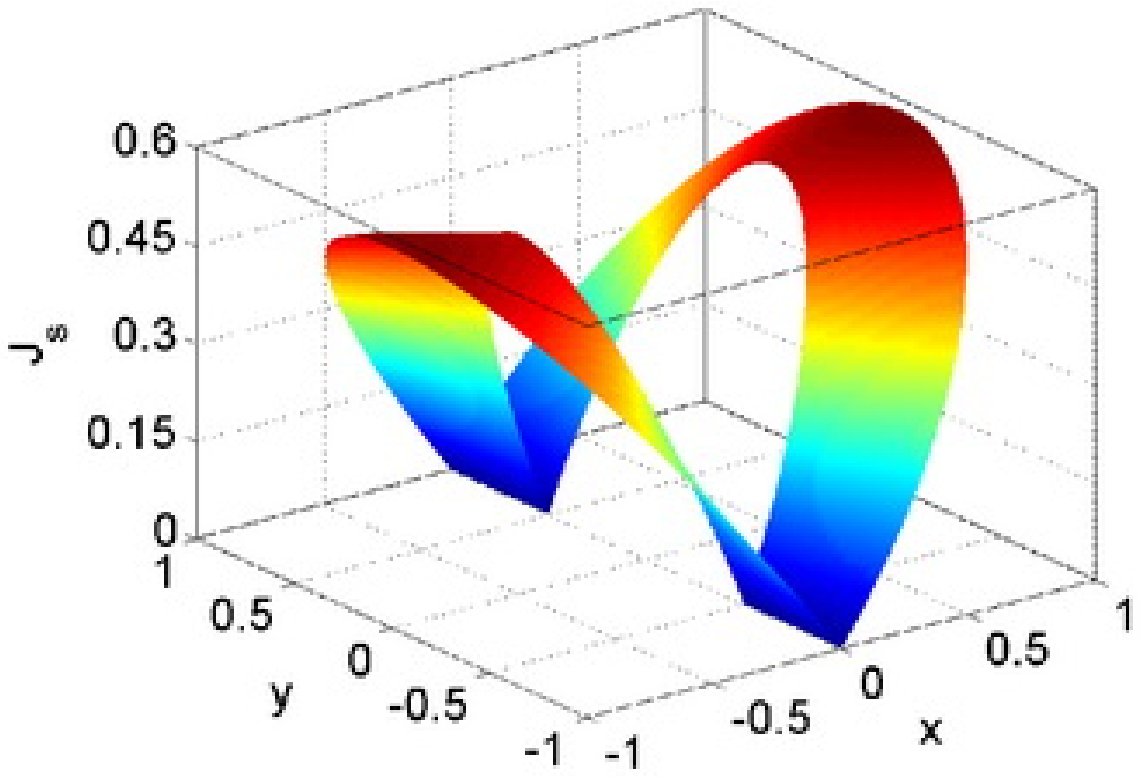}}
\epsfysize=3.0cm \put(0.0,0.0){\epsffile{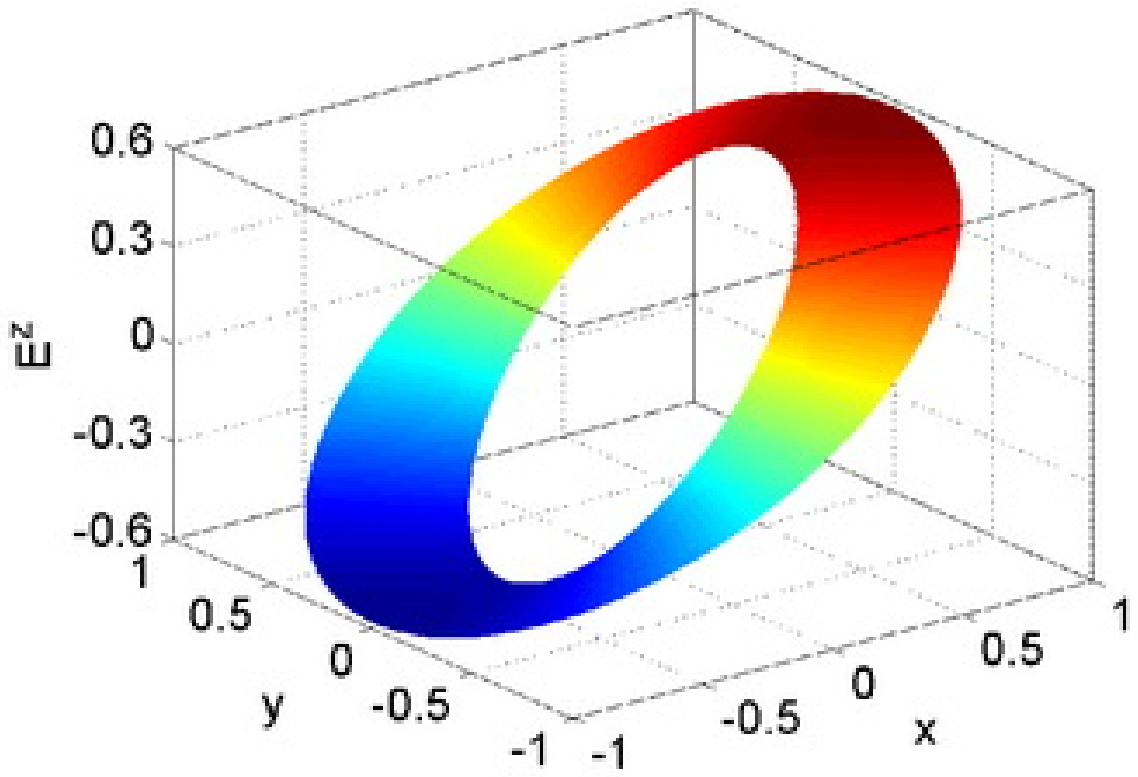}}
\epsfysize=3.0cm \put(4.5,0.0){\epsffile{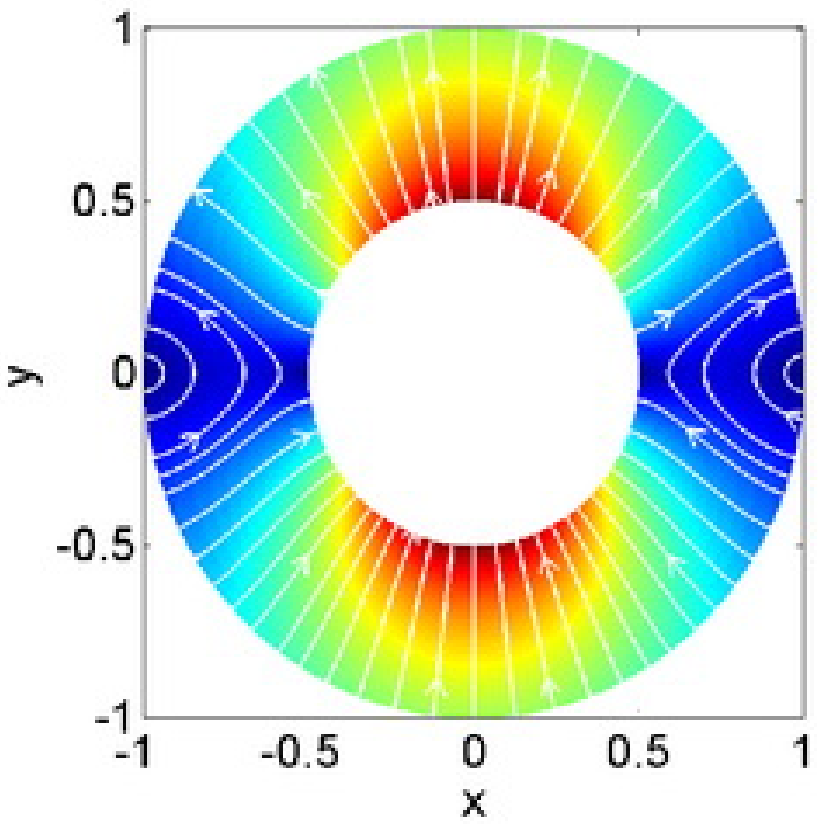}}
\put(0.2,5.8){(a)} \put(4.2,5.8){(b)} \put(0.2,2.8){(c)}
\put(4.2,2.8){(d)}
\end{picture}
\caption{(color online). Same as Fig.~\ref{cm11} for the (1,1) mode
of an annular mesa, except that the lateral coordinates are
normalized by the outer radius and the ratio between the inner and
outer radii is 1/2.} \label{am11}
\end{figure}

The distributions of the static phase kink, supercurrent, the
spatial part of the oscillating electric and magnetic field are
displayed in Fig.~\ref{am11} for the case of $a_{\rm i}/a_{\rm
o}=1/2$, with $k^{a}_{11}=\chi^{a}_{11}/a_{\rm o}=1.3546/a_{\rm o}$
from Eq.~(\ref{vam11}).

There are two features concerning the redistribution of magnetic
field, comparing Fig.~\ref{am11}(d) and Fig.~\ref{cm11}(d). First,
the maximum of magnetic field at the cylinder center is suppressed.
Secondly, the magnetic field becomes along the radial direction in
most part of the mesa when the center of mesa is removed, since the
magnetic field should be normal to both outer and inner surfaces.
Both of these two factors suppress the strength of magnetic field,
and thus reduce the frequency.

\begin{figure}
\epsfysize=8cm \epsfclipoff \fboxsep=0pt
\setlength{\unitlength}{1cm}
\begin{picture}(8,6)(0,0)
\epsfysize=3.0cm \put(0.0,3.0){\epsffile{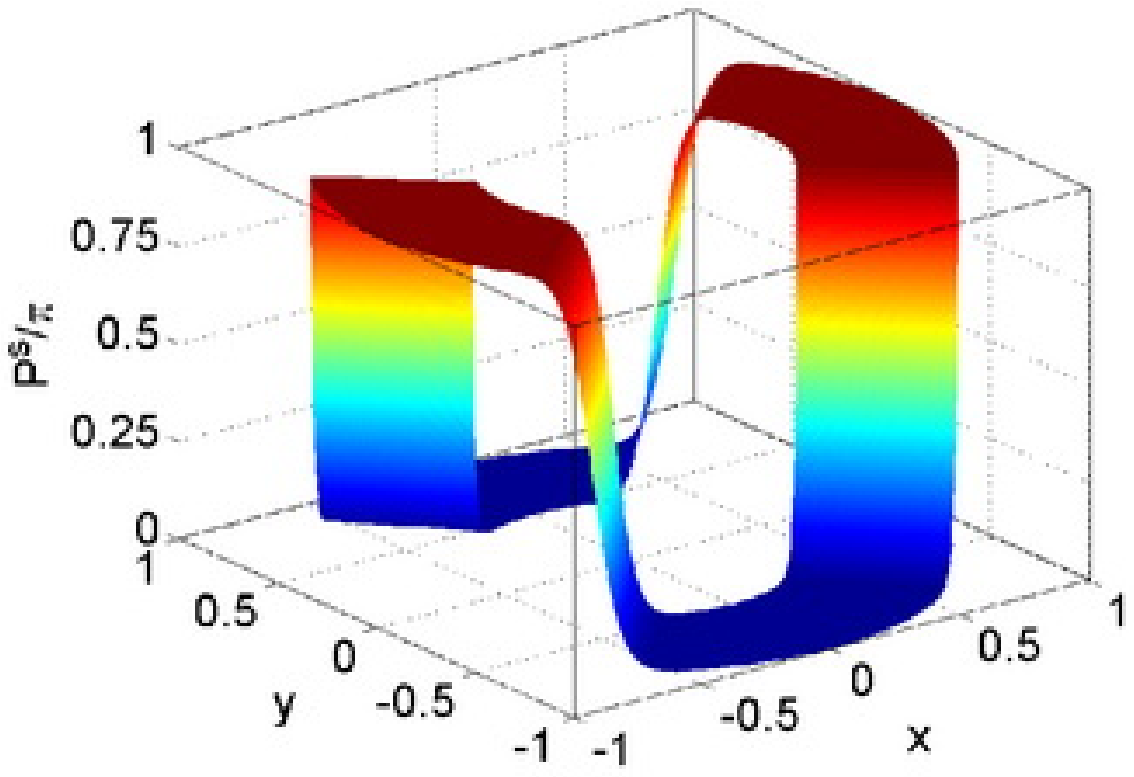}}
\epsfysize=3.0cm \put(4.0,3.0){\epsffile{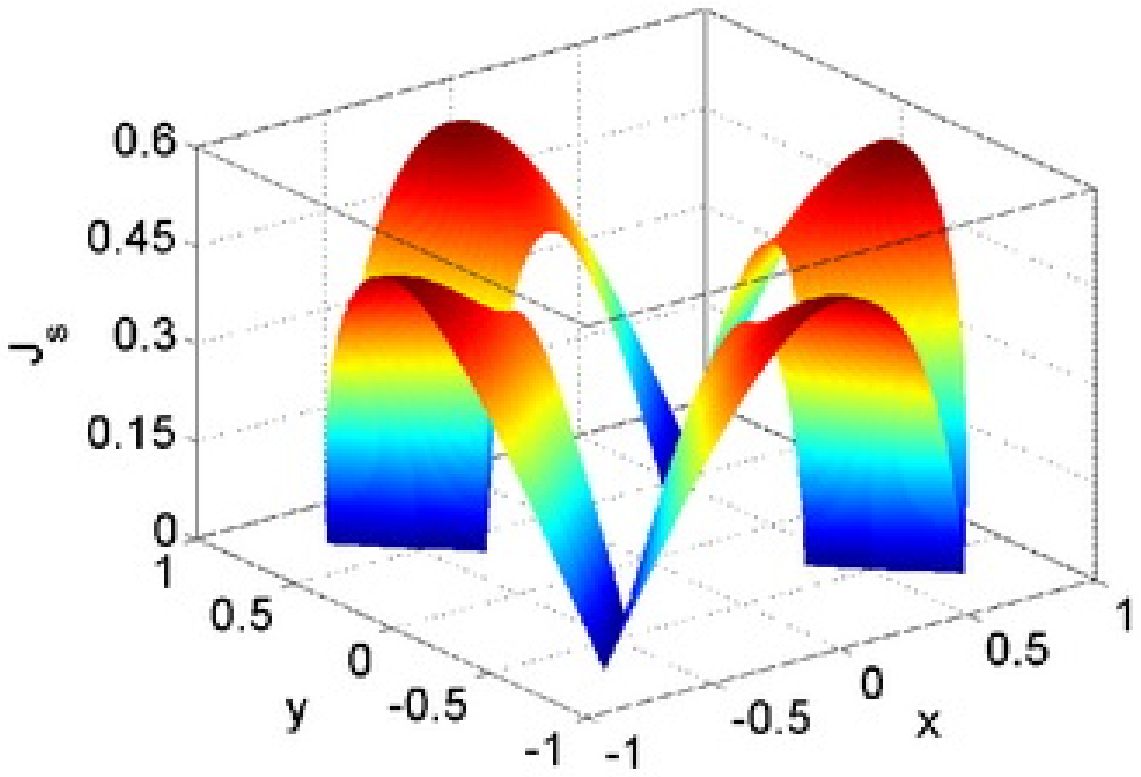}}
\epsfysize=3.0cm \put(0.0,0.0){\epsffile{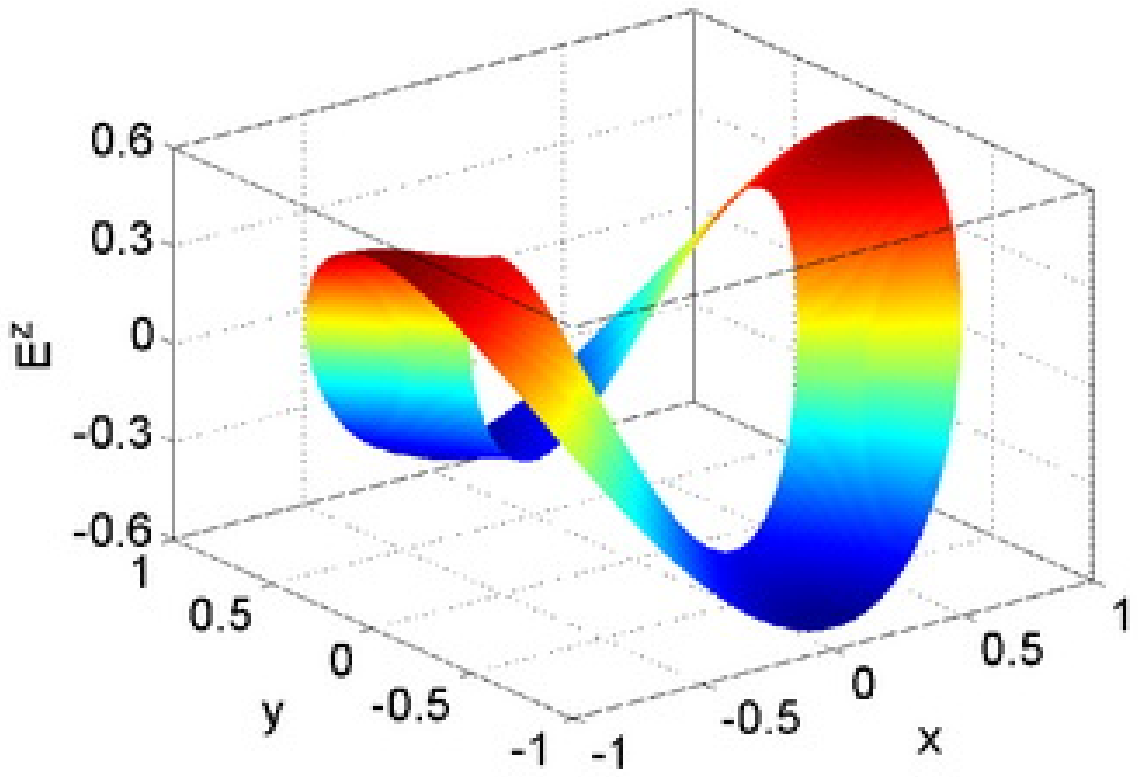}}
\epsfysize=3.0cm \put(4.5,0.0){\epsffile{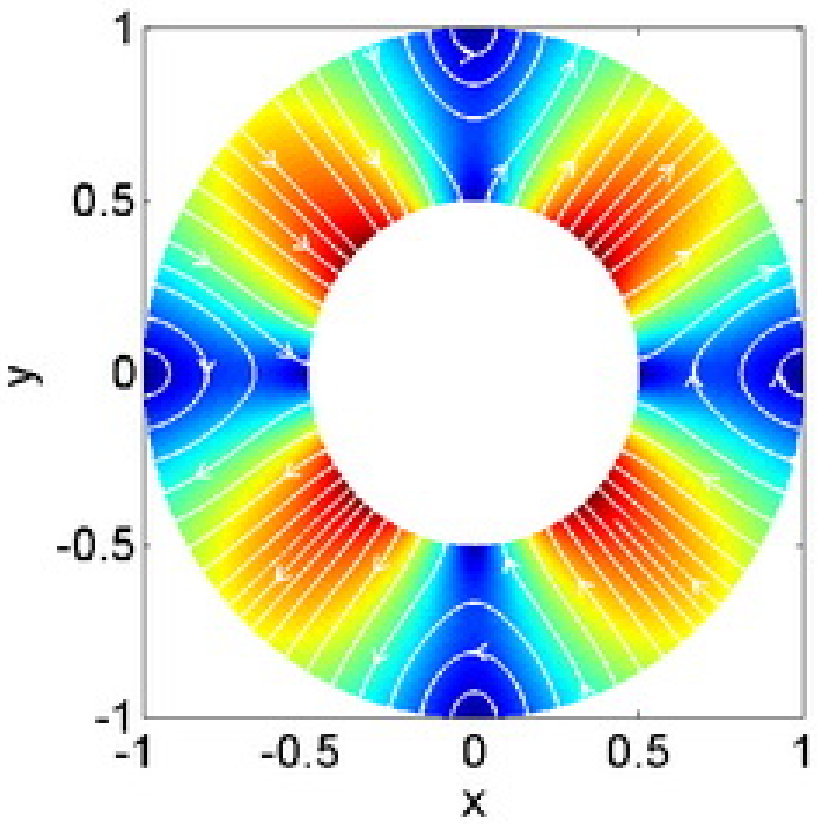}}
\put(0.2,5.8){(a)} \put(4.2,5.8){(b)} \put(0.2,2.8){(c)}
\put(4.2,2.8){(d)}
\end{picture}
\caption{(color online). Same as Fig.~\ref{am11} for the (2,1) mode
of an annular mesa.} \label{am21}
\end{figure}

\subsection{(2,1) mode for annular mesa}

The plasma term for (2,1) mode is given by

\begin{equation}
\Tilde{P}(\bm{r},t)= \Bigl(A_{\rm J} J_2(\frac{\chi^{a}_{21}}{a_{\rm
o}}\rho) + A_{\rm N} N_2(\frac{\chi^{a}_{21}}{a_{\rm o}}\rho )\Bigr)
\cos(2\phi) \sin(\omega t+\varphi). \label{phaseam21}
\end{equation}

\noindent The wave number of this mode is determined by an equation
same to Eq.~(\ref{vam11}) except for using $J'_2(z)$ and $N'_2(z)$
instead of $J'_1(z)$ and $N'_1(z)$ as
$k^{a}_{21}=\chi^{a}_{21}/a_{\rm o}=2.6812/a_{\rm o}$ for $a_{\rm
i}/a_{\rm o}=1/2$. The distributions of the static phase kink,
supercurrent, the spatial part of the oscillating electric and
magnetic field are displayed in Fig.~\ref{am21}, which is to be
compared with Fig.~\ref{cm21}. Similar to (1,1) mode, the magnetic
field becomes along the radial direction in most part of the mesa.

\begin{figure}
\epsfysize=8cm \epsfclipoff \fboxsep=0pt
\setlength{\unitlength}{1cm}
\begin{picture}(8,6)(0,0)
\epsfysize=3.0cm \put(0.0,3.0){\epsffile{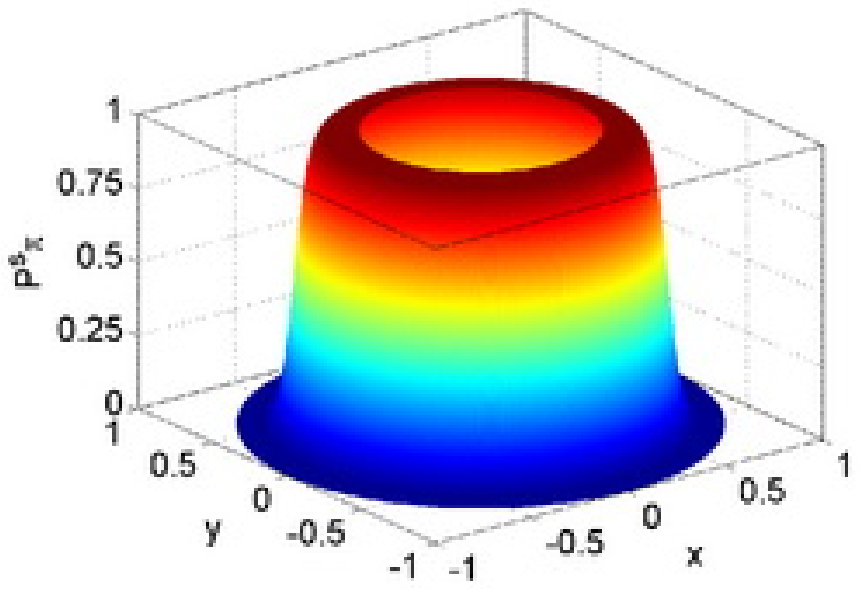}}
\epsfysize=3.0cm \put(4.0,3.0){\epsffile{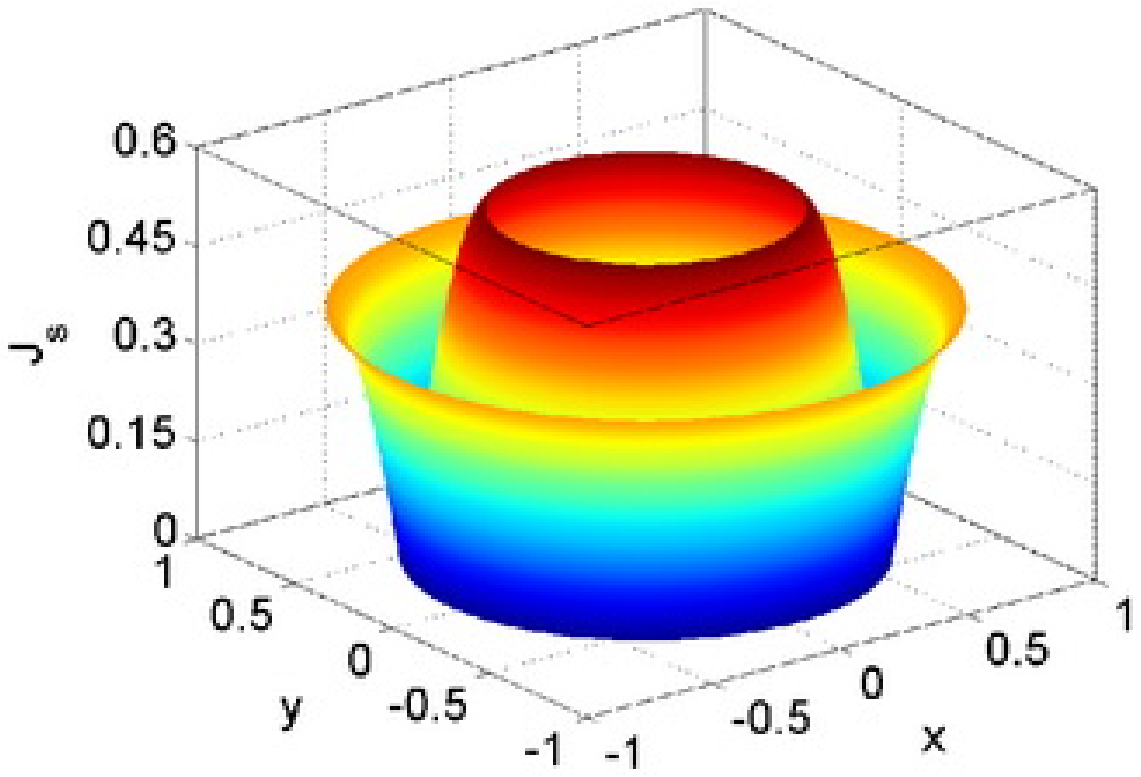}}
\epsfysize=3.0cm \put(0.0,0.0){\epsffile{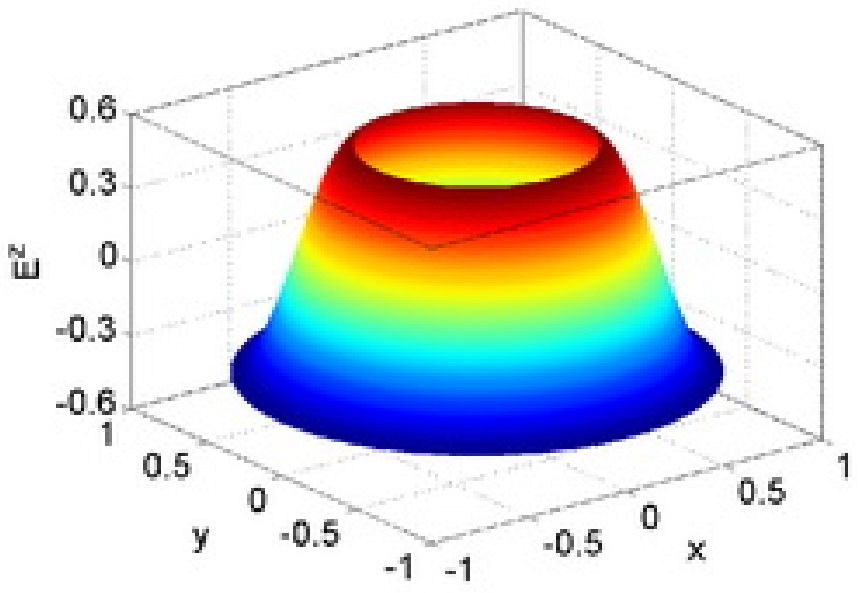}}
\epsfysize=3.0cm \put(4.5,0.0){\epsffile{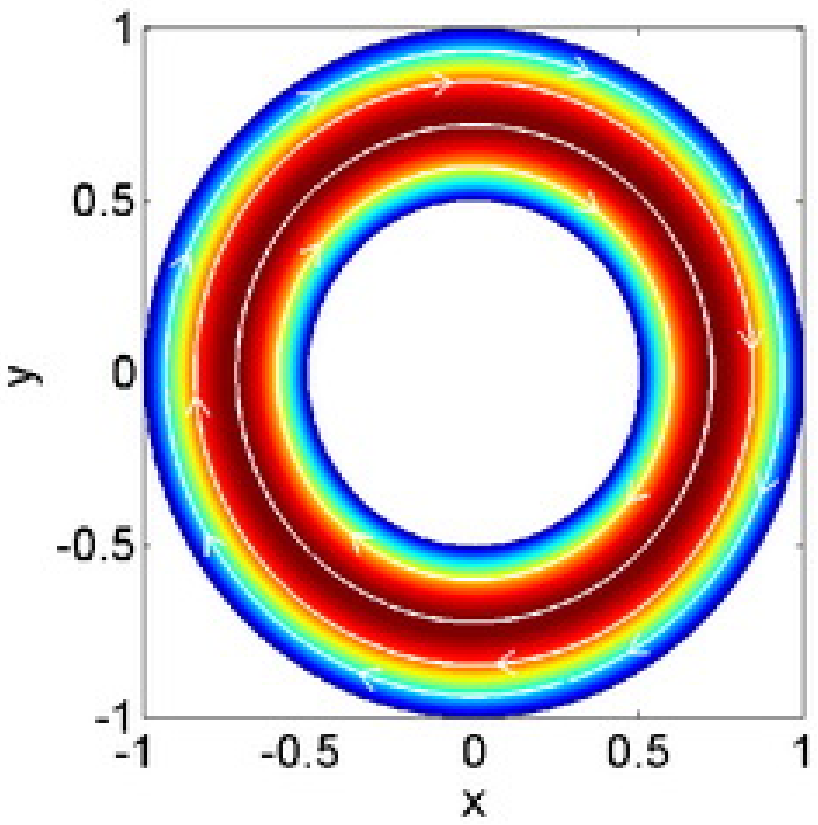}}
\put(0.2,5.8){(a)} \put(4.2,5.8){(b)} \put(0.2,2.8){(c)}
\put(4.2,2.8){(d)}
\end{picture}
\caption{(color online). Same as Fig.~\ref{am11} for the (0,1) mode
of an annular mesa.} \label{am01}
\end{figure}

\begin{figure}
\epsfysize=8cm \epsfclipoff \fboxsep=0pt
\setlength{\unitlength}{1cm}
\begin{picture}(8,6)(0,0)
\epsfysize=3.0cm \put(0.0,3.0){\epsffile{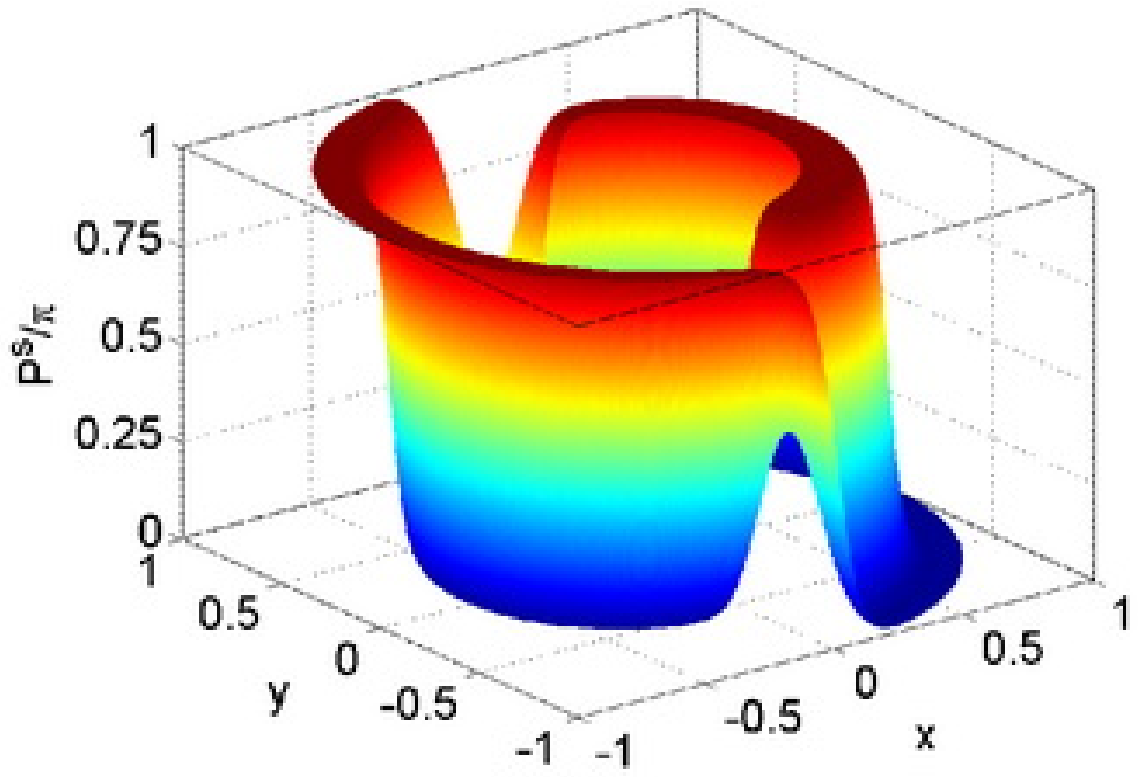}}
\epsfysize=3.0cm \put(4.0,3.0){\epsffile{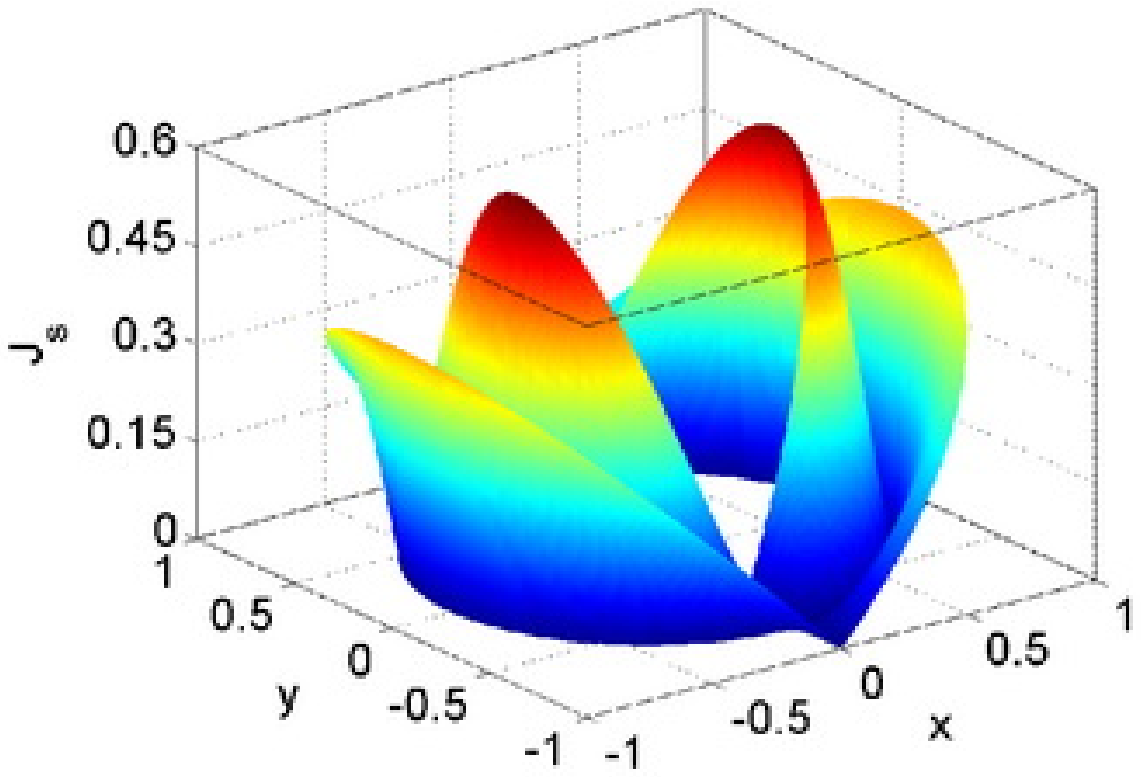}}
\epsfysize=3.0cm \put(0.0,0.0){\epsffile{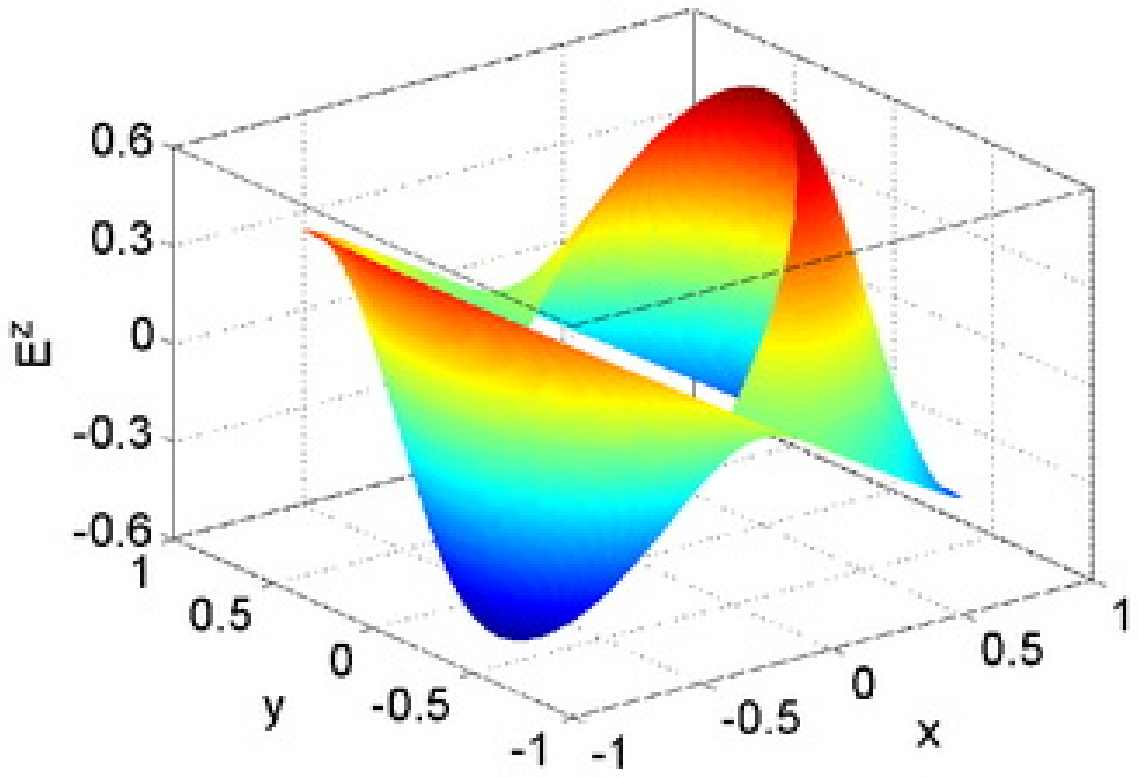}}
\epsfysize=3.0cm \put(4.5,0.0){\epsffile{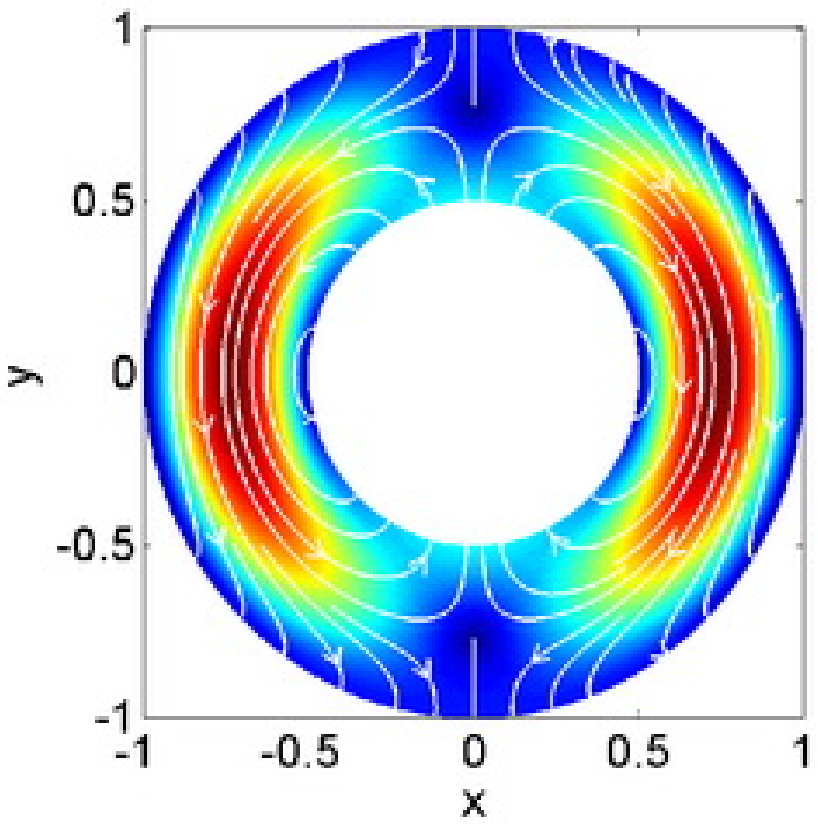}}
\put(0.2,5.8){(a)} \put(4.2,5.8){(b)} \put(0.2,2.8){(c)}
\put(4.2,2.8){(d)}
\end{picture}
\caption{(color online). Same as Fig.~\ref{am11} for the (1,2) mode
of an annular mesa.} \label{am12}
\end{figure}

\subsection{(0,1) mode for annular mesa}

The plasma term for the (0,1) mode is

\begin{equation}
\Tilde{P}(\bm{r},t)= \Bigl(A_{\rm J} J_0(\frac{\chi^{a}_{01}}{a_{\rm
o}}\rho) + A_{\rm N} N_0(\frac{\chi^{a}_{01}}{a_{\rm o}}\rho )\Bigr)
\sin(\omega t+\varphi). \label{phaseam01}
\end{equation}

\noindent The wave number is $k^{a}_{01}=\chi^{a}_{01}/a_{\rm
o}=6.3932/a_{\rm o}$ for $a_{\rm i}/a_{\rm o}=1/2$. The
distributions of the static phase kink, supercurrent, the spatial
part of the oscillating electric and magnetic field are displayed in
Fig.~\ref{am01}.

In contrast to the modes (1,1) and (2,1), the magnetic field is
circular in the (0,1) mode. The system arranges the magnetic field
to zero totally at the two surfaces to satisfy the Newmann boundary
condition.  Thus, the density of magnetic flux increases when the
center of cylinder is removed, which results in a higher frequency.

\subsection{(1,2) mode for annular mesa}

The plasma term for the (1,2) mode is

\begin{equation}
\Tilde{P}(\bm{r},t)= \Bigl(A_{\rm J} J_1(\frac{\chi^{a}_{12}}{a_{\rm
o}}\rho) + A_{\rm N} N_1(\frac{\chi^{a}_{12}}{a_{\rm o}}\rho
\Bigr)\cos\phi \sin(\omega t+\varphi). \label{phaseam12}
\end{equation}

\noindent The wave number is $k^{a}_{12}=\chi^{a}_{12}/a_{\rm
o}=6.5649/a_{\rm o}$ for $a_{\rm i}/a_{\rm o}=1/2$. The
distributions of the static phase kink, supercurrent, the spatial
part of the oscillating electric and magnetic field are displayed in
Fig.~\ref{am12}.

There are two features in the redistribution of magnetic field for
this mode as displayed in Fig.~\ref{am12}(d). The maximum of
magnetic field at the center of the cylindrical mesa disappears (
Fig.~\ref{cm12}(d)), similar to (1,1) mode. On the other hand, the
density of magnetic flux increases at the other two maxima of
magnetic field, similar to (0,1) mode.

\begin{figure}[t]
\setlength{\unitlength}{1cm}
 \psfig{figure=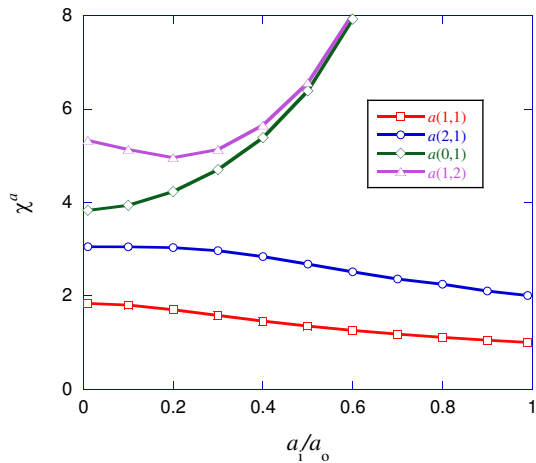,width=7cm}
\caption{(color online). Wave-number factor $\chi^{a}$ as a function
of the radius ratio $a_{\rm i}/a_{\rm o}$ for the lowest four modes
for the annular geometry, with the wave number given by
$k^a=\chi^a/a_{\rm o}$.} \label{vamfig}
\end{figure}

\subsection{Dependence of frequency on aspect ratio}

The dependence of wave number, and equivalently frequency, on the
aspect ratio $a_{\rm i}/a_{\rm o}$ is shown in Fig.~\ref{vamfig}.
For the (1,1) and (2,1) modes, the wave number $k^a=\chi^a/a_{\rm
o}$ decreases with increasing ratio $a_{\rm i}/a_{\rm o}$, whereas
an opposite trend is seen for the (0,1) mode. The behavior of the
(1,2) mode is a compromise of the both trends, yielding a minimum in
wave number at $a_{\rm i}/a_{\rm o}\simeq 0.2$. These behaviors are
well understood from the redistribution of magnetic field in the
corresponding cavity modes discussed above.

For the (1,2) mode, the regime $0\le a_{\rm i}/a_{\rm o}\le 0.4$ can
be used to suppress the Joule heating without changing much the
radiation frequency. Since the curves for the (1,1) and (2,1) modes
in Fig.~\ref{vamfig} are quite flat, they can also be useful for the
same purpose. The mode (0,1) and the mode (1,2) with $a_{\rm
i}/a_{\rm o}>0.4$ can be used to increase the radiation frequency
provided the heating effect can be controlled.

\begin{figure}[t]
\setlength{\unitlength}{1cm}
 \psfig{figure=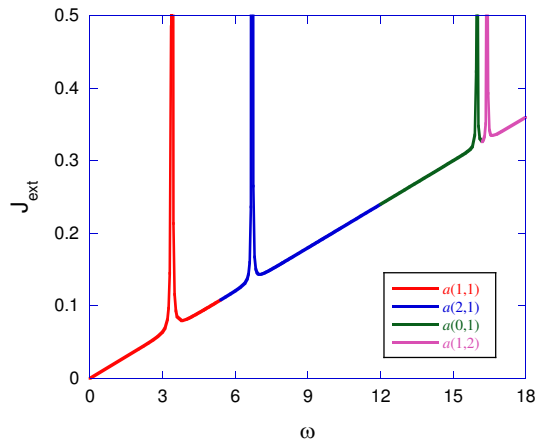,width=7cm}
\caption{(color online). \emph{IV} characteristics for the annular
mesa including the four lowest modes. The dimensionless voltage is
$\omega=\chi^a/a$ with $\chi^a=1.3546$, 2.6812, 6.3932 and 6.5649
for the (1,1), (2,1), (0,1) and (1,2) modes respectively. The radii
are $a_{\rm i}=0.2$ and $a_{\rm o}=0.4$.} \label{ivam}
\end{figure}

\subsection{\emph{IV} characteristics for annular mesa}

The \emph{IV} characteristics for the annular mesa is also given by
Eq.~(\ref{iv}). The prefactor for the (1,1), (2,1), (0,1) and (1,2)
mode is 0.405, 0.403, 0.396 and 0.321, respectively. The \emph{IV}
characteristics for the annular mesa is displayed in
Fig.~\ref{ivam}.

\subsection{Radiation pattern for annular mesa}

\begin{figure}
\epsfysize=8cm \epsfclipoff \fboxsep=0pt
\setlength{\unitlength}{1cm}
\begin{picture}(8,6.8)(0,0)
\epsfysize=3.5cm \put(0.0,3.5){\epsffile{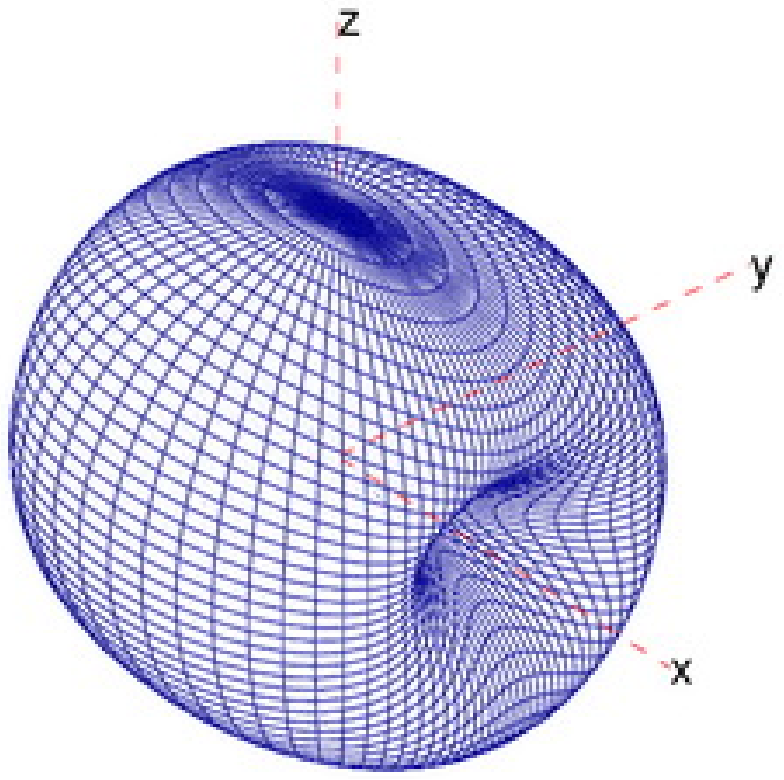}}
\epsfysize=4.0cm \put(4.0,3.2){\epsffile{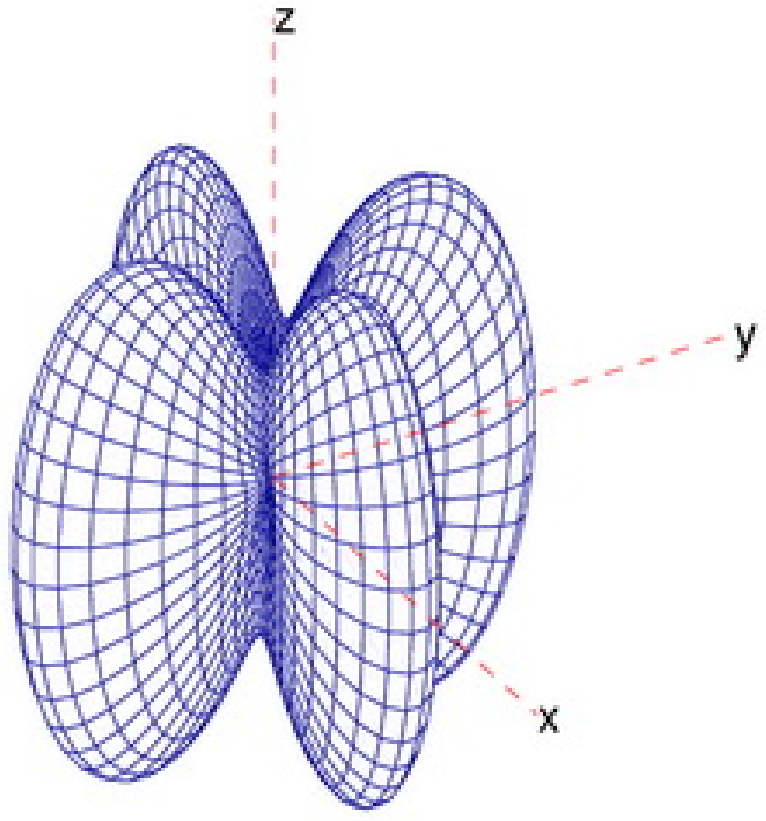}}
\epsfysize=3.5cm \put(0.0,0.0){\epsffile{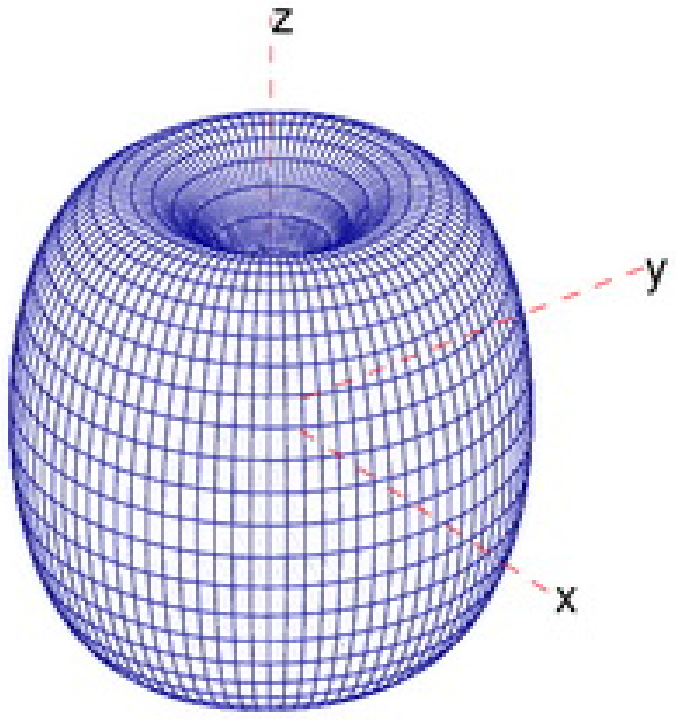}}
\epsfysize=3.5cm \put(4.0,0.0){\epsffile{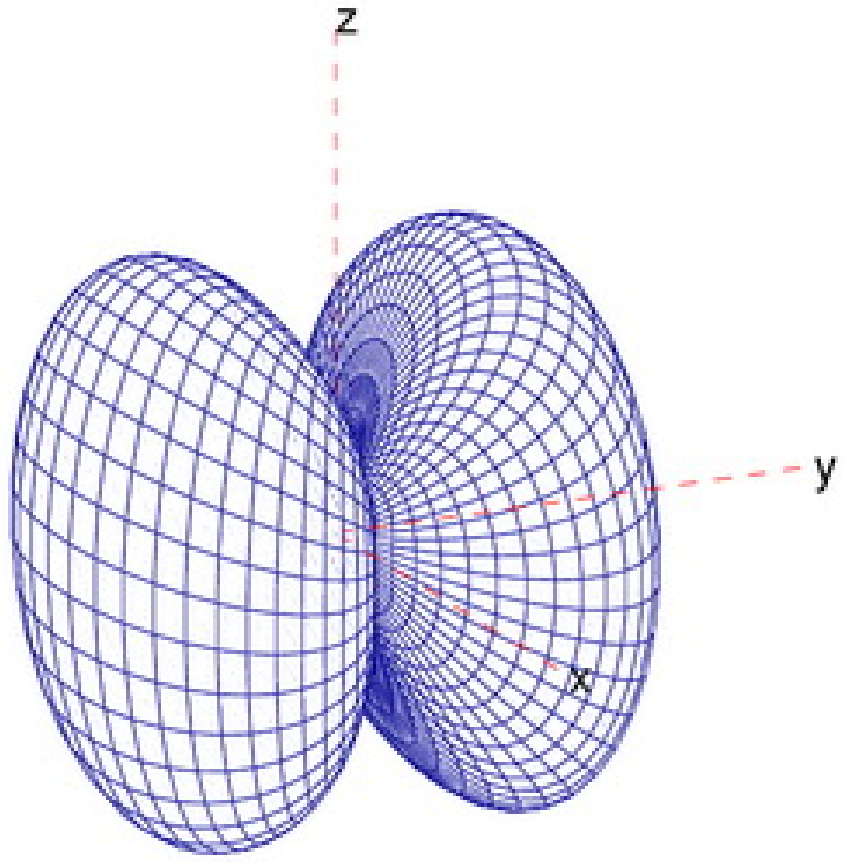}}
\put(0.2,6.6){(a)} \put(4.2,6.6){(b)} \put(0.2,3.1){(c)}
\put(4.2,3.1){(d)}
\end{picture}
\caption{(color online). Same as Fig.~\ref{patterncm} for annular
mesa. Multi reflections due to the inner surface are neglected since
the thickness of annular mesa is very small compared with the EM
wavelength. The radii are $a_{\rm i}=0.2$ and $a_{\rm o}=0.4$.}
  \label{patternam}
\end{figure}

 The radiation pattern of each mode is displayed in
Fig.~\ref{patternam} for the annular geometry, where multi
reflections at the inner radius have been neglected since the mesa
thickness is very small compared with the EM wavelength.

\section{Summary and Perspectives}

We first address that the radius dependence of radiation frequency
observed in a recent experiment using a cylindrical mesa
\cite{Kadowaki09} is to be understood as a clear indication of the
right boundary condition for mesa of small thickness compared with
the EM wavelength, with which the static $\pi$ kink state has been
derived uniquely for inductively coupled Josephson junctions.
Detailed spatial distributions of the superconductivity phase
difference with a $\pi$ phase kink are presented for various cavity
modes of cylindrical mesa. Along with them, we also summarize the
spatial distribution of the EM standing waves inside the junctions
which hopefully can be observed in experiments.

We propose to use annular mesas to excite THz EM radiations. The
obvious advantage of the annular geometry is the reduction of
heating, since the area of sample is reduced which suppresses the
total Joule heating and the inner surface of the sample may enhance
heat leakage additionally. The effect of removing the central part
of a cylindrical mesa, thus rendering an annular one, on the
redistribution of the superconductivity phase difference, the
supercurrent and the EM waves is analyzed. It is shown that,
depending on the mode, the radiation frequency varies with the
aspect ratio in different ways, and there is plenty of room for
modification of the EM radiation by tailoring the shape of sample,
while keeping low heating.

Superconductor samples of annular geometry can also be manufactured
as a waveguide resonator, where fragments of superconductor mesas
are arranged in a thread with each satisfying the condition of small
thickness compared with the wavelength of electromagnetic wave. In
contrast with the fiber lasers developed for visible lights (see for
example \cite{laser}), the lasing mode propagates along the annular
superconductor, whereas the electromagnetic waves within the inner
surface establishes the coherent oscillation in thread of
superconductor mesas. When the outer surface is shielded by wrapping
a jacket made of appropriate materials, the wave propagating in the
wave guide can be enhanced.

\section{Acknowledgements}
This work was supported by WPI Initiative on Materials
Nanoarchitectonics, MEXT of Japan and by CREST, JST of Japan, and
partially by ITSNEM of CAS.


\begin{thebibliography}{99}
\bibitem{Note} In most literatures, the first zero the derivative of Bessel
function $J_0(z)$ is put as 3.8317. The better description is to put
0 as the first zero and 3.8317 as the second, and so on so forth.
The mode shown in Fig.~\ref{cm01} (see also Ref.\cite{HuLinPRB08})
would then be assigned as the (0,2) mode in cylinder. The redefined
mode (0,1) is the uniform solution, which is not a real cavity mode
\cite{LinHuPRB09}.
\end{thebibliography}
\end{document}